\documentclass[fleqn,usenatbib]{mnras}

\usepackage[T1]{fontenc}

\DeclareRobustCommand{\VAN}[3]{#2}
\let\VANthebibliography\thebibliography
\def\thebibliography{\DeclareRobustCommand{\VAN}[3]{##3}\VANthebibliography}

\usepackage{graphicx}	% Including figure files
\usepackage{amsmath}	% Advanced maths commands
\usepackage{amssymb}	% Extra maths symbols

\usepackage{newtxtext,newtxmath}
\usepackage{subfigure}
\usepackage{color}
\usepackage{epsfig}
\usepackage{array}
\usepackage{url}
\usepackage{bm}
\title[Assessment of the cosmic distance duality relation using Gaussian Process]{Assessment of the cosmic distance duality relation using Gaussian Process}

\author[Purba Mukherjee, Ankan Mukherjee]{
Purba Mukherjee$^{1}$\thanks{E-mail: pm14ip011@iiserkol.ac.in}
and Ankan Mukherjee$^{2,3}$\thanks{Email: ankan.ju@gmail.com}
\\
$^{1}$Department of Physical Sciences, Indian Institute of Science Education and Research Kolkata, Mohanpur 741246, India\\
$^{2}$Department of Physics, Bangabasi College, Kolkata 700009, India\\
$^{3}$Centre for Theoretical Physics, Jamia Millia Islamia, New Delhi 110025, India.
}

\date{Accepted ??. Received ??; in original form ??}

\pubyear{????}

\begin{document}
\label{firstpage}
\pagerange{\pageref{firstpage}--\pageref{lastpage}}
\maketitle

\begin{abstract}
Two types of distance measurement are important in cosmological observations, the angular diameter distance $d_A$ and the luminosity distance $d_L$. In the present work, we carried out an assessment of the theoretical relation between these two distance measurements, namely the cosmic distance duality relation, from type Ia supernovae (SN-Ia) data, the Cosmic Chronometer (CC) Hubble parameter data, and baryon acoustic oscillation (BAO) data using Gaussian Process. The luminosity distance curve and the angular diameter distance curve are extracted from the SN-Ia data and the combination of BAO and CC data respectively using the Gaussian Process. The distance duality relation is checked by a non-parametric reconstruction using the reconstructed $H$, $d_L$, and the volume-averaged distance $D_v$. We compare the results obtained for different choices of the covariance function employed in the Gaussian Process. It is observed that the theoretical distance duality relation is in well agreement with the present analysis in 2$\sigma$ for the overlapping redshift domain $0 \leq z \leq 2$ of the reconstruction.
\end{abstract}


%%%%%%%%%%%%%%%%%%%%%%%%%%%%%%%%%%%%%%%%%%%%%%%%%%

%%%%%%%%%%%%%%%%% BODY OF PAPER %%%%%%%%%%%%%%%%%%

\section{Introduction}

Distance measurement in cosmology gives an idea about the mutual separation between two objects or events in the universe on radial null trajectories which terminate at the observer. It serves as one of the most crucial and important tasks involved in cosmography, helping us to establish a standard relation between the observational data with theoretical models, thus setting up the rudimentary framework. Distance measures are often used to tie some observable quantity (such as the luminosity of a distant star, or angular size of the acoustic peaks in the Cosmic Microwave Background power spectrum) to some other quantity that is not directly observable, but is more convenient for calculations.\\

Cosmologists use different definitions of \textit{distance} from the observer to a celestial object at redshift $z$, of which two are extremely useful. They are the luminosity distance $d_L$, and the angular diameter distance $d_A$ respectively. The former is defined by the relation between bolometric flux and bolometric luminosity, which can further be connected to the absolute magnitude $M$ and apparent magnitude $m$ of an astronomical object. Likewise, the latter one is a distance associated  with an object's physical size, and it's angular size, projected on the celestial sphere. They are connected through the cosmic distance-duality relation (CDDR) \citep{etherin1993, etherin2007} given as,
\begin{equation} \label{ccdr}
d_L = d_A (1+z)^{2}.
\end{equation} 
The CDDR was first proved by \citeauthor{etherin1993} in the context of a Friedmann-Lema\^{i}tre-Robertson-Walker (FLRW) metric, and is often recognized as Etherington's reciprocity theorem.\\

The reciprocity theorem is considered to be correct for any general metric theories of gravity in any background based on two fundamental hypotheses. Firstly, when  number of photons is conserved during cosmic evolution \citep{ellis1971, ellis2007}, and secondly, when gravity is described by a metric theory with photons travelling on unique null geodesics in a Riemannian geometry. The coupling of photons with unknown particles \citep{bassett}, the extinction of photons by intergalactic dust \citep{corasaniti}, the variation of fundamental constants \citep{ellis2013} can lead towards a violation of CDDR. As cosmography is strongly dependent on the validity of CDDR, a little deviation from it may indicate the possibility of some exotic physics beyond the standard model or the presence of systematic errors in observations \citep{bassett,bassett2}. Therefore, with the increasing quality and quantity of observational data, evaluating the reliability of CDDR has received much attention lately.\\

The CDDR can straight away be put to test utilizing the luminosity distance $d_L (z)$ and the angular diameter distance $d_A (z)$ measurements at the same redshift $z$. The distance measures can generally be obtained through observations of the type Ia supernovae (SN-Ia), high redshift galaxies (HII), angular diameter of galaxy clusters, galaxy cluster gas mass fraction, strong gravitational lensing (SGL), cosmic microwave background (CMB), gamma ray bursts (GRB), radio compact sources, baryon acoustic oscillations (BAO), Gravitational Waves (GWs), and so on. The type-Ia supernovae data \citep{union2.1,betoule2014,pan1} generally serve as major sources  for estimating $d_L$ with high precision. Nevertheless, determining the angular diameter distance $d_A$ is neither simple nor effortless as that of $d_L$. The combined data of the X-ray and Sunyaev-Zeldovich (SZ) effect of galaxy clusters \citep{filipps2005,bonamente2006}, the baryon acoustic oscillations (BAO) in the galaxy power spectrum \citep{beutler2011,blake2012,anderson2014,kazin2014}, the angular size of ultra-compact radio sources based on the approximately consistent linear size \citep{kellermann1993,gurvits1994,gurvits1999,jackson2004}, the images of quasars that are strongly gravitational lensed by foreground galaxies \citep{cao2015,liao2016} are amongst a few feasible measurement practices. However, computing $d_A$ as discussed above have their respective merit and limitations. \citet{li2018, lin2018} provides us with a brief discussion in this context.\\

While testing for validity of CDDR, the prime difficulty experienced is that both $d_L$ and $d_A$ are not computed at the same redshift $z$. Several approaches have been proposed to resolve this drawback. These involves the nearest neighbourhood method \citep{holanda2010, liao2016}, the interpolation method \citep{liang2013}, and the Gaussian Processes \citep{nair3, rana2017, li2018} method. It deserves mention that only data points in the overlapping redshift range are available for CDDR verification. Until now, no evidence for CDDR violation has been recorded from the reconstruction techniques. In the nearest neighbour approach \citet{holanda2010} considered two sub-samples of SN-Ia from the Constitution data along with two samples of galaxy clusters compiled by \citeauthor{filipps2005} and \citeauthor{bonamente2006} by combining the SZ effect and X-ray surface brightness.  The SN-Ia redshifts of each sub-sample by were carefully chosen within a redshift separation $\Delta z < 0.005$, so that they coincide with those of the associated galaxy cluster samples, thereby allowing a direct test for the CDDR. However, this approach suffers from several inconsistencies, viz. significant reduction of statistical information encoded in the data sets as pointed out by \citet{max2014}. This redshift-matching problem from the nearest neighbour method was overcome by \citet{cardone2012}, applying a local regression technique to the SN-Ia $d_L$ data at the concerned redshift windows with adjustable bandwidth. But again, this idea could not be easily generalized to strongly correlated data, thus rejecting a majority of data points, thereby leading to incorrect estimation. \citet{ma2016, ma2018} addressed this issue by using Bayesian statistical techniques, precisely a Monte Carlo method, which compresses correlated luminosity distance data at specific points in log-redshift. \\

Reconstruction of the CCCR has previously been often addressed in literature. They are categorized into the cosmological model-dependent tests \citep{uzan2004,avgous2010,holanda2011,lima2011,nair,jhingan2014,holanda2016b,hu2018}, and the cosmological model-independent analysis \citep{holanda2010,li2011,goncalves2011,nair2,meng2012,chen2015,holanda2016,liao2016,rana2016,holanda2017,holanda2019,xu2020,zheng2020}. \citet{nair} studied the validity of six different parametrized CDDR using the Union2 SN data. \citet{holanda2012} used the gas mass fraction measurements of galaxy clusters from SZ effect and X-ray surface brightness observations. \citet{gonclaves2014} considered the gas mass fraction measurements reported by the Atacama Cosmology Telescope (ACT) survey along with Union2.1 SN-Ia compilation. \citet{costa2015} used galaxy clusters and Hubble data measurements. \citet{holanda2016} utilized the nearest neighbour approach considering SGL and JLA SN-Ia weighted average data. \citet{liao2016} considered a compilation of SGL and JLA SN-Ia, along with galaxy cluster samples and FRIIb radio galaxies. \citet{rana2016} worked with the JLA SN-Ia data, and samples from radio galaxies. \citet{hu2018} tested CDDR in the $R_h = ct$ Universe using the regression method. \citet{li2018} used the ultra-compact radio sources in combination with Union 2.1 SN-Ia data. \citet{ma2018} used the BAO data from BOSS DR12 and WriggleZ survey, along with JLA SN-Ia compilation. \citet{ruan2018} considered SGL and HII galaxy Hubble diagram to obtain constraints on the CDDR parametrizations. Testing CDDR from Future GWs Sirens was done by \citet{fu2019}. However, all these works involve a functional form for the CDDR chosen at the outset, followed by an estimation of parameters. This is undoubtedly biased, as a specific parametric form for the CDDR is already chosen.\\

In the present work, the Supernova distance modulus data, Cosmic Chronometer measurements of the observational Hubble data and the Baryon Acoustic Oscillation data have been utilized in examining the validity of CDDR in a non-parametric way. The reconstruction is performed adopting the model-independent Gaussian Processes (GP) method. For a detailed overview one can refer to the Gaussian Process website\footnote{\url{http://www.gaussianprocess.org}}. A non-parametric reconstruction of CDDR using the GP method is barely addressed in the existing literature. \citet{nair3} compared distance measurements from the Union2 sample of supernovae with BAO data from SDSS, 6dFGS and the latest BOSS and WiggleZ surveys. \citet{rana2017} presented a new way to constrain the CDDR using different dynamic and geometric properties of SGL along with JLA SN-Ia observations. Results showed no violation of CDDR. Here, we use the recent Pantheon sample of SN-Ia for obtaining $d_L$, and $d_A$ are derived considering the volume-averaged BAO compilation data in combination with the CC $H(z)$ data, at the same domain of redshift. We attempt to perform this reconstruction avoiding any fiducial bias on the cosmological parameters included in the data sets. \\

This paper is arranged as follows. Section \ref{reconst} contains the details on the reconstruction method. In section \ref{datasets}, the observational data used in the present work have been briefly reviewed.  The results obtained are presented in section \ref{results}. We conclude the manuscript in section \ref{conclusion} with an overall discussion about the results.

\section{Reconstruction Methodology}
\label{reconst}

In this section we shall discuss the details of the reconstruction technique and define the necessary physical quantities involved in our work. It is already mentioned that an entirely non-parametric, model-independent approach is utilized in the present study. A reconstruction of the cosmic distance duality relation has been carried out directly from the observational data without assuming any specific parametrization, or an \textit{a priori} fiducial background cosmology. We utilize the Gaussian Process method as our numerical tool in this work.\\

To test the validity of the cosmic distance duality relation, we analyse the following redshift dependence of CDDR, given by, 
\begin{equation} \label{eta}
\eta(z) = \frac{d_L}{d_A(1+z)^2}.
\end{equation}
In case of $\eta = 1$ we recover back equation \eqref{ccdr}, which implies a non-violation of the standard CDDR. Any deviation of $\eta$ from unity ($\eta \neq 1$) indicates a non-validation of CDDR. In order to test the CDDR directly from observations, we perform a model independent GP reconstruction of $\eta$ using the publicly available \texttt{GaPP}\footnote{\url{https://github.com/carlosandrepaes/GaPP}} (Gaussian Processes in Python) code developed by \citet{gapp}.\\

The uncertainty associated with the reconstructed function $\eta$ can be calculated by the standard error propagation formula, 
\begin{equation} \label{sig_eta}
\sigma_\eta = \eta \sqrt{\left( \frac{\sigma_{d_L}}{d_L}\right)^2 + \left( \frac{\sigma_{d_A}}{d_A} \right)^2} ,
\end{equation}

The Gaussian Process is a generalization of the Gaussian probability distribution. It generalizes the idea of a Gaussian distribution characterized by discrete data points to the continuous limit. Assuming the observational data, say $D$, obeys a Gaussian distribution with mean and variance, the posterior distribution of the target function $\bm{F}={F(z_i)}_{i=1}^n$ describing the data $\left\lbrace(z_i, D_i)\vert i=1,\cdots n\right\rbrace$ can be expressed via the joint Gaussian distribution of $D(z_i)$'s. Thus, given a set of Gaussian-distributed data points one can use GP to reconstruct the target function, and also obtain the associated confidence levels, without assuming a concrete parametrization of the target function. For this the key element is a covariance function $k(z, \tilde{z})$ which correlates values of the reconstructed function $D$ at redshift points $z$ and $\tilde{z}$ separated by $\vert z-\tilde{z} \vert$ distance units. This covariance function $k(z, \tilde{z})$ depends on a set of hyperparameters (i.e., the characteristic length scale $l$ and the signal variance $\sigma_f$). $l$ describes the distance one needs to move roughly in input space before the function value changes significantly, whereas $\sigma_f$ determines a distinct change in the function value. Different choices for the covariance function may have different effects on the reconstruction. Here, we take into account the Squared Exponential as well as three orders of the Mat\'{e}rn  covariance functions in performing our analysis. The Squared Exponential covariance function is defined as,
\begin{equation} \label{sqexp}
k(z, \tilde{z}) = \sigma_f^2 \exp \left( - \frac{(z-\tilde{z})^2}{2l^2}\right)
\end{equation}
and the Mat\'{e}rn covariance is given by,
\begin{equation} \label{matern}
\begin{split}
k_{\nu=p+\frac{1}{2}}(z,\tilde{z}) = \sigma_f^2 \exp \left( \frac{-\sqrt{2p+1}}{l} \vert z - \tilde{z} \vert \right) \times \\ \times \frac{p!}{(2p)!} \sum_{i=0}^{p} \frac{(p+i)!}{i!(p-i)!} \left( \frac{2\sqrt{2p+1}}{l} \vert z - \tilde{z} \vert \right)^{p-i} .
\end{split}
\end{equation}
In cases for the Mat\'{e}rn covariance function, the orders of the polynomials are taken as $(p=2,3,4)$, consequently the parameter $\nu$ has the values $(\frac{5}{2},\frac{7}{2},\frac{9}{2})$. \\

In order to implement the Gaussian Process and reconstruct the function $\bm{F}$, one needs to know the hyperparameters $\sigma_f$ and $l$. They can be trained by maximizing the marginal likelihood, which is a marginalization over function values $\{F(z_i)\}$ at observational redshift locations $\{z_i\}_{i=1}^n$. Note that this marginal likelihood is independent of the redshift points where we want to reconstruct the function. For a Gaussian prior the log marginal likelihood is given by, \begin{equation}
\begin{split}
\ln \mathcal{L} = -\frac{1}{2}(\bm{D-\mu})^{\mbox{\small T}} \left[ K(\bm{Z}, \bm{Z}) + \bm{\mathcal{C}} \right]^{-1}(\bm{D-\mu}) + \\ -\frac{1}{2} \ln\vert K(\bm{Z}, \bm{Z})+\bm{\mathcal{C}} \vert -\frac{n}{2}\ln 2\pi
\end{split}
\end{equation} where, $\bm{\mu}= \{\mu(z_i)\}_{i=1}^n$ is the mean function and $K(\bm{Z},\bm{Z})$ is the covariance matrix given by $[K(\bm{Z},\bm{Z})]_{ij} = k(z_i, z_j)$ at $\bm{Z}= \{z_i\}_{i=1}^n$ observational redshift points. $\bm{\mathcal{C}}$ is the covariance matrix of the data and $n$ is the dimension of $\bm{D}$ or $\bm{Z}$.

\begin{table*} 
	\caption{{\small Table showing the reconstructed value of $H_0$ (in units of km Mpc$^{-1}$ s$^{-1}$) for different choices of the covariance function, from the CC data.}}
	\begin{center}
		\resizebox{0.98\textwidth}{!}{\renewcommand{\arraystretch}{1.3} \setlength{\tabcolsep}{20pt} \centering  
			\begin{tabular}{c c c c c} 
				%				\hline
				\hline
				$k(z,\tilde{z})$ & Mat\'{e}rn 9/2 & Mat\'{e}rn 7/2 & Mat\'{e}rn 5/2 &  Squared Exponential\\ 
				\hline
				\hline
				$H_0$  &  $68.471 \pm 5.081$  & $68.684 \pm 5.204$ & $68.858 \pm 5.466$	& $67.356 \pm 4.765$\\ 
				\hline
			\end{tabular}
		}
	\end{center}
	\label{Hz_res}
\end{table*}

%%%%%%%%%%%%%%%%%%%%%
\begin{figure*}%
	\begin{center}
		\includegraphics[angle=0, width=0.245\textwidth]{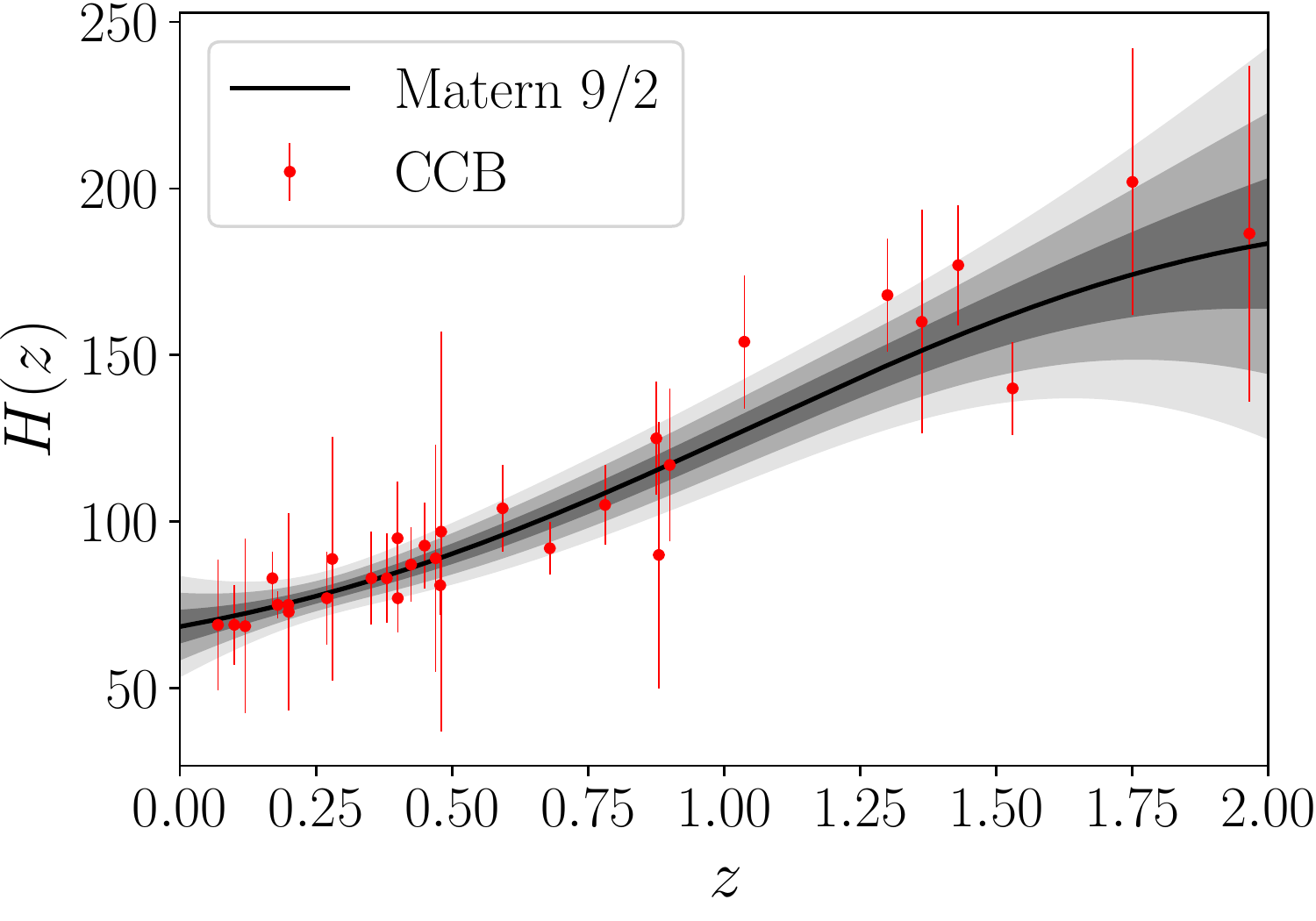}
		\includegraphics[angle=0, width=0.245\textwidth]{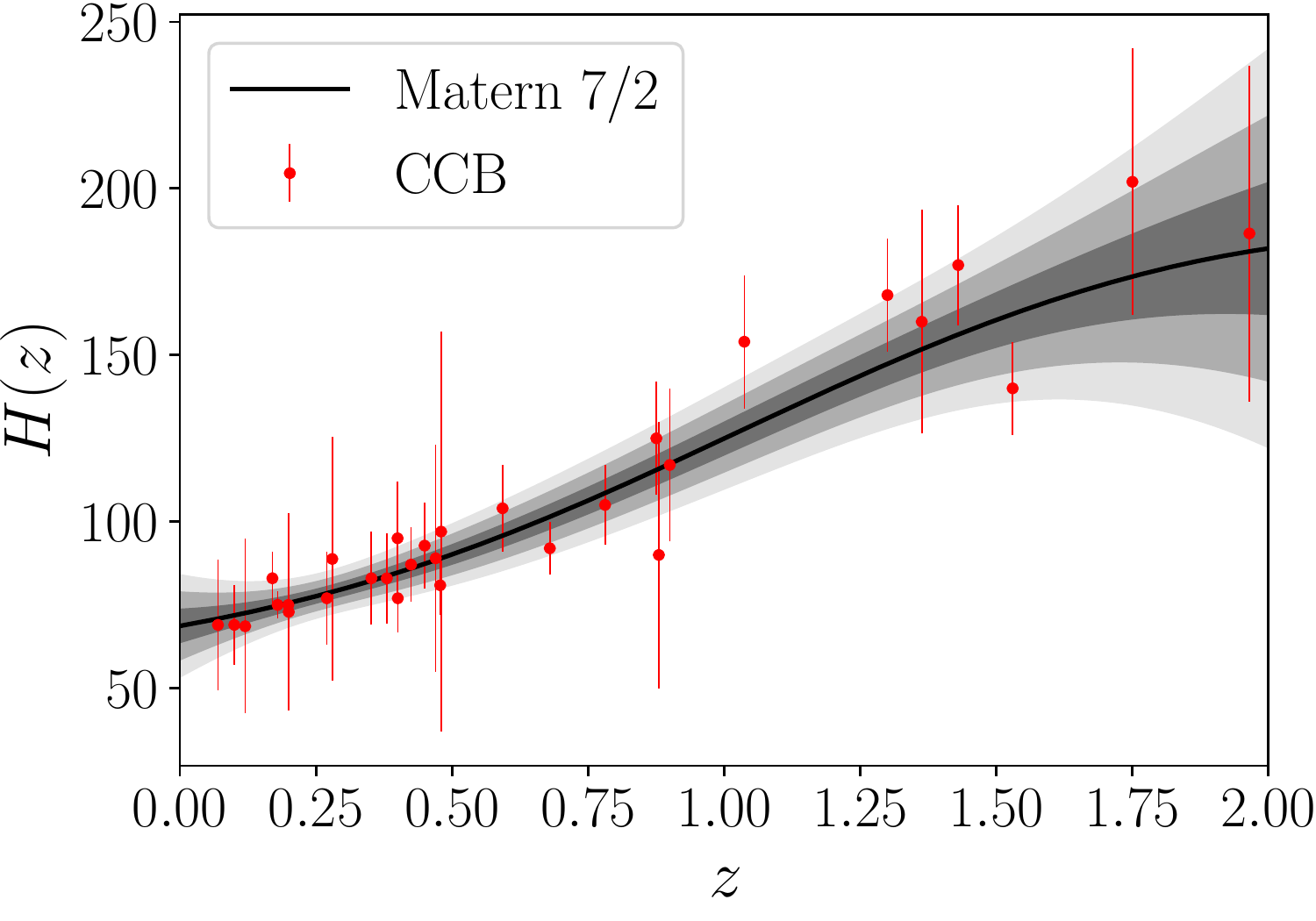}
		\includegraphics[angle=0, width=0.245\textwidth]{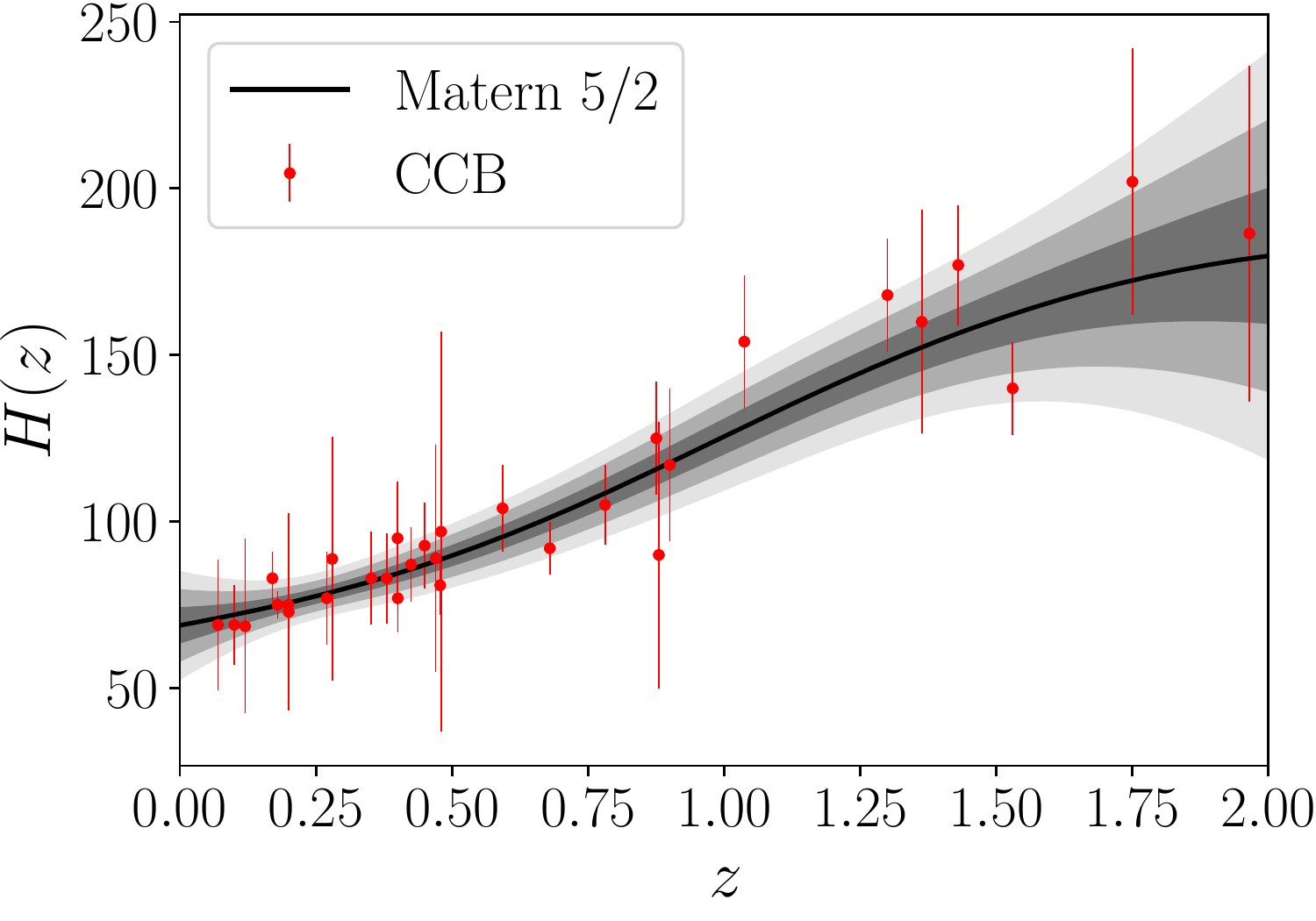}
		\includegraphics[angle=0, width=0.245\textwidth]{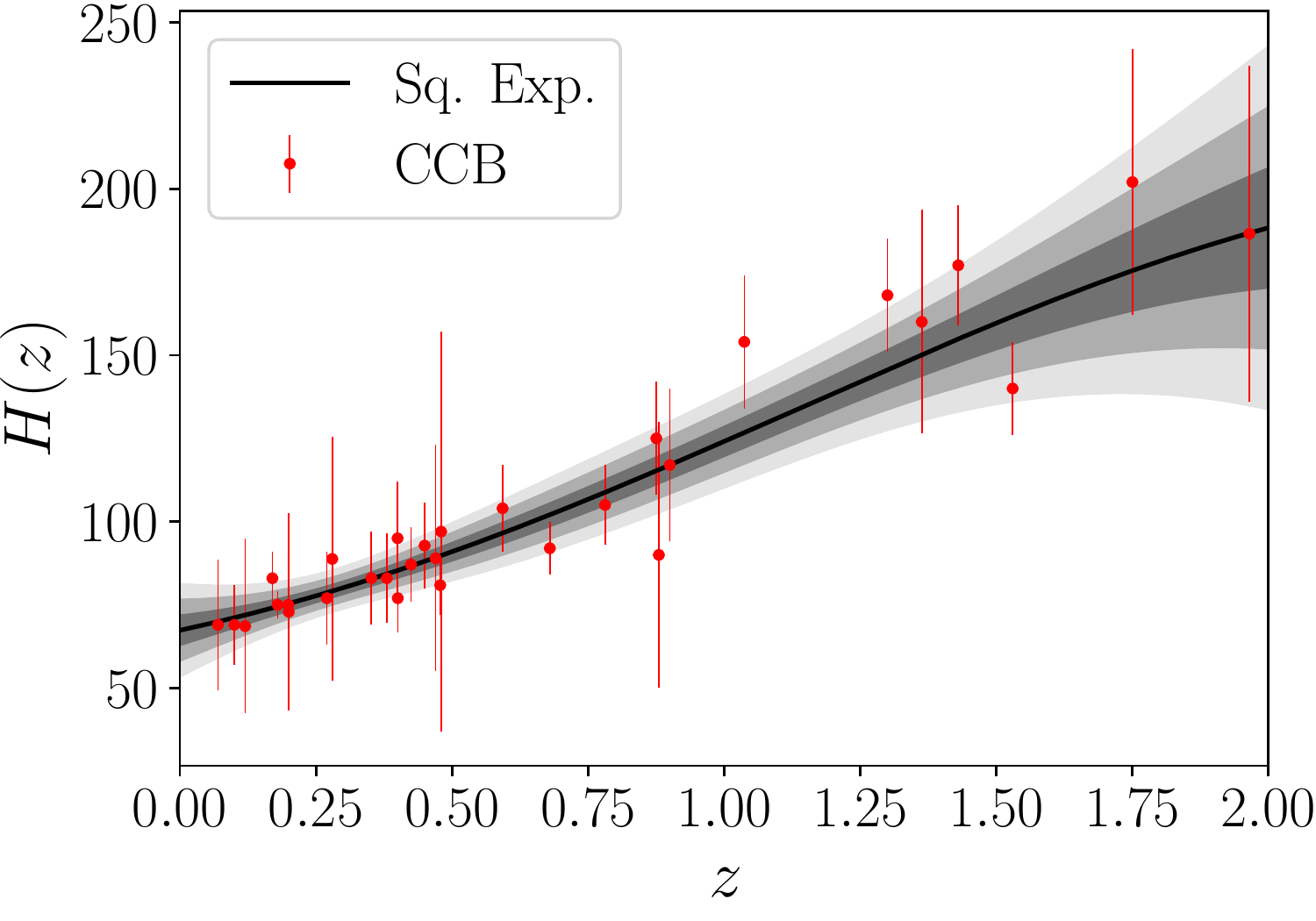}
	\end{center}
	\caption{{\small Plots for $H(z)$ (in units of km Mpc$^{-1}$ s$^{-1}$) reconstructed from CC data using the Mat\'{e}rn 9/2, Mat\'{e}rn 7/2, Mat\'{e}rn 5/2 and Squared Exponential covariance function (from left to right) respectively. The solid black line is the best fitting curve and the associated 1$\sigma$, 2$\sigma$ and 3$\sigma$ confidence regions are shown in lighter shades.}}
	\label{Hz_recon}
\end{figure*}

%%%%%%%%%%%%%%%%%%%%%
\begin{table*}
	\caption{{\small Table showing the marginalized constraints on $M_B$, $\Omega_{m0}$ and $r_d$ (in units of Mpc) for different choices of the covariance function using equations \eqref{chiSN}, \eqref{chiCC} and \eqref{chiBAO} respectively.}}
	\begin{center}
		\resizebox{0.98\textwidth}{!}{\renewcommand{\arraystretch}{1.3} \setlength{\tabcolsep}{20pt} \centering  
			\begin{tabular}{c c c c c} 
				%				\hline
				\hline
				$k(z,\tilde{z})$  &  Mat\'{e}rn 9/2 & Mat\'{e}rn 7/2 & Mat\'{e}rn 5/2 & Squared Exponential\\ 
				\hline
				\hline
				$M_B$ & $-19.388 \pm 0.006$ &  $-19.387 \pm 0.006$  & $-19.387 \pm 0.006$ & $-19.391 \pm 0.006$	\\ 
				\hline
				$\Omega_{m0}$ & $0.305 \pm 0.004$ &  $0.302 \pm 0.004 $  & $0.299 \pm 0.005$ & $0.321 \pm 0.004 $	\\ 
				\hline
				$r_d$ & $146.116 \pm 0.336$ &  $146.086 \pm 0.340$  & $146.044 \pm 0.350 $ & $146.193 \pm 0.326$	\\ 
				\hline
			\end{tabular}
		}
	\end{center}
	\label{MOr_res}
\end{table*}
%%%%%%%%%%%%%%%%%%%%%

\section{Observational Datasets}
\label{datasets}

It is already mentioned that the SN-Ia distance modulus data, the Cosmic Chronometer measurements of the Hubble parameter and the BAO volume-averaged distance data are adopted in the present context. In the following section, these data sets and the way they are utilized in the present analysis are discussed.

\subsection{$d_L(z)$ from SN-Ia data}

The type Ia supernovae are widely accepted as {\it standard candles} to measure cosmological distance for their approximately consistent absolute luminosity \citep{riess1998, perl1999}. For the supernova data, we use the recent Pantheon compilation sample by \citet{pan1} consisting of 1048 SN-Ia data points. The numerical data of the full Pantheon SN-Ia catalogue is publicly available\footnote{\url{http://dx.doi.org/10.17909/T95Q4X}}$^,$\footnote{\url{https://archive.stsci.edu/prepds/ps1cosmo/index.html}}.\\

The Pantheon compilation is at present the largest sample which consists of different supernovae surveys, including SDSS, SNLS, various LOW-$z$ samples and some HIGH-$z$ samples from HST. We include the covariance matrix along with systematic errors in our calculation. The distance modulus of each supernova is given by,
\begin{equation}\label{mu}
\mu(z) = 5 \log_{10} \left(\frac{d_L(z)}{\mbox{Mpc}}\right) + 25
\end{equation}
where $d_L$ is the luminosity distance. So, \begin{equation}
d_L(z)  = 10^{\left( \frac{\mu}{5} - 5 \right)}
\end{equation} in units of Mpc.\\

Now, the distance modulus of SN-Ia can be derived from the observation of light curves through the empirical relation,
\begin{equation}
\mu_{\mbox{\tiny SN}} = m^{*}_B + \alpha X_1 - \beta C - M_B + \Delta_M + \Delta_B
\end{equation} where $X_1$ and $C$ are the stretch and colour parameters, and $M_B$ is the absolute magnitude  in the B-band for SN-Ia. $\alpha$ and $\beta$ are two nuisance parameters, $\Delta_M$ is a distance correction based on the host-galaxy mass of the SN-Ia and $\Delta_B$ is a distance correction based on predicted biases from simulations. Usually, the nuisance parameters $\alpha$ and $\beta$ are optimized simultaneously with the cosmological model parameters or are marginalized over. However, this method is model dependent. In the Pantheon sample, the corrected apparent magnitude $m_B = m^{*}_B + \alpha X_1 - \beta C$ along with $\Delta_M$ and $\Delta_B$ corrections are reported  following the BEAMS with Bias Corrections [BBC] \citep{beams} framework. Therefore, the colour, stretch corrections are no longer required, so we can fix $\alpha = \beta = 0$ and proceed with $\mu_{\mbox{\tiny SN}} = (m_B - M_B)$. \\

\begin{figure*}%
	\begin{center}
		\includegraphics[angle=0, width=0.2\textwidth]{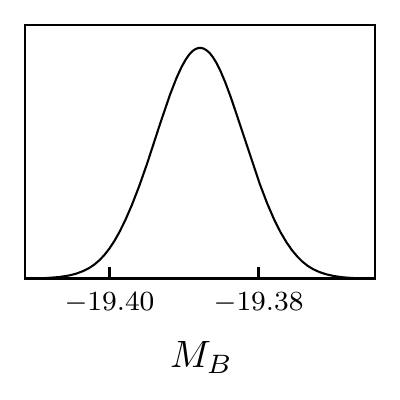}~~~~~~~~~~~~
		\includegraphics[angle=0, width=0.2\textwidth]{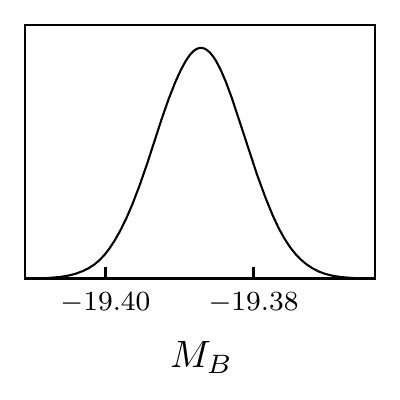}~~~~~~~~~~~~
		\includegraphics[angle=0, width=0.2\textwidth]{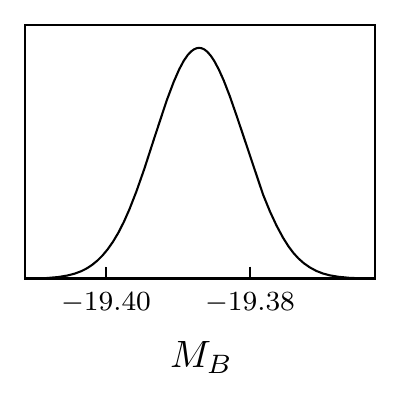}~~~~~~~~~~~~
		\includegraphics[angle=0, width=0.2\textwidth]{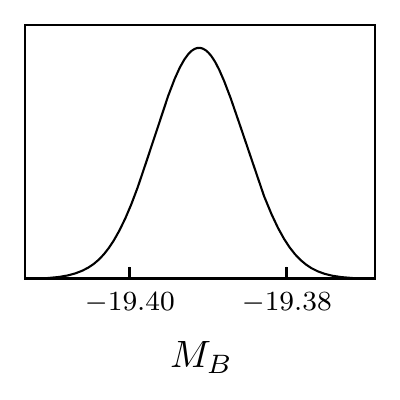}
	\end{center}
	\caption{{\small Plots for marginalized likelihood of absolute magnitude $M_B$ using the Mat\'{e}rn 9/2, Mat\'{e}rn 7/2, Mat\'{e}rn 5/2 and Squared Exponential covariance function (from left to right) respectively.}}
	\label{MB_plot}
\end{figure*}
%%%%%%%%%%%%%%%%%%%%%

\begin{figure*}%
	\begin{center}
		\includegraphics[angle=0, width=0.2\textwidth]{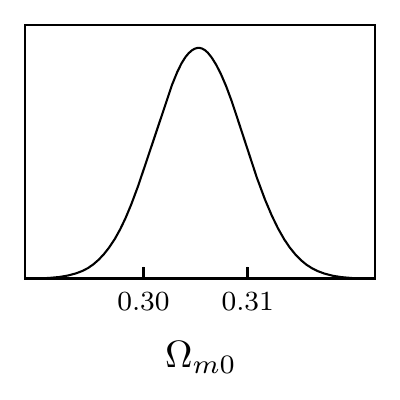}~~~~~~~~~~~~
		\includegraphics[angle=0, width=0.2\textwidth]{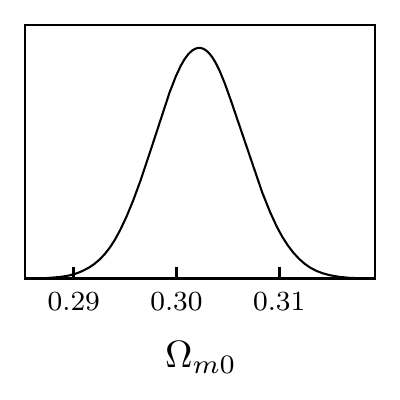}~~~~~~~~~~~~
		\includegraphics[angle=0, width=0.2\textwidth]{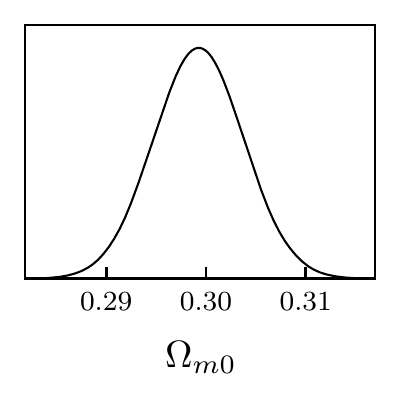}~~~~~~~~~~~~
		\includegraphics[angle=0, width=0.2\textwidth]{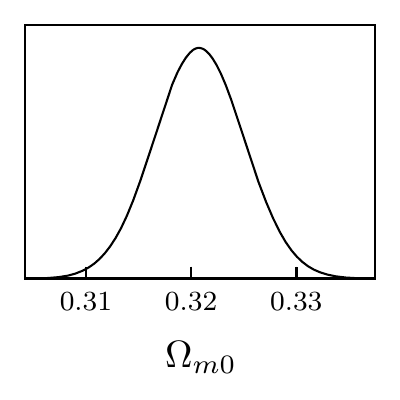}
	\end{center}
	\caption{{\small Plots for marginalized likelihood of matter density parameter  $\Omega_{m0}$ using the Mat\'{e}rn 9/2, Mat\'{e}rn 7/2, Mat\'{e}rn 5/2 and Squared Exponential covariance function (from left to right) respectively.}}
	\label{Om_plot}
\end{figure*}
%%%%%%%%%%%%%%%%%%%%%

\begin{figure*}%
	\begin{center}
		\includegraphics[angle=0, width=0.2\textwidth]{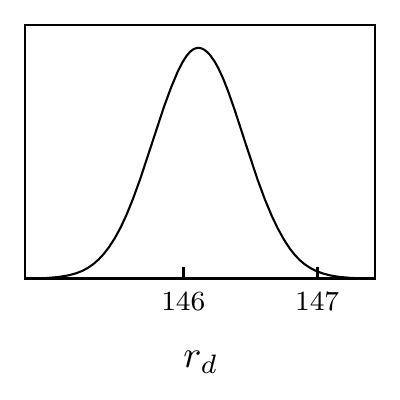}~~~~~~~~~~~~
		\includegraphics[angle=0, width=0.2\textwidth]{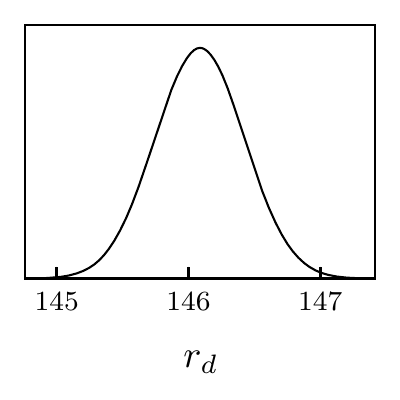}~~~~~~~~~~~~
		\includegraphics[angle=0, width=0.2\textwidth]{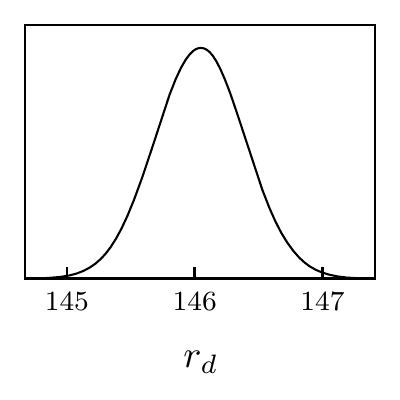}~~~~~~~~~~~~
		\includegraphics[angle=0, width=0.2\textwidth]{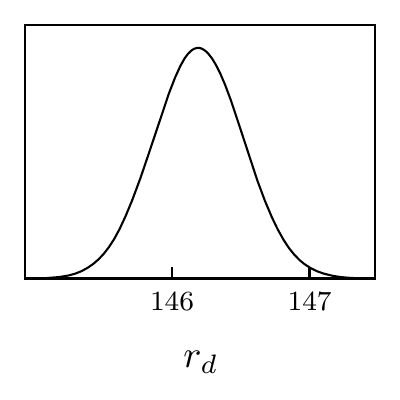}
	\end{center}
	\caption{{\small Plots for marginalized likelihood of comoving sound horizon at drag epoch $r_d$ (in units of Mpc) using the Mat\'{e}rn 9/2, Mat\'{e}rn 7/2, Mat\'{e}rn 5/2 and Squared Exponential covariance function (from left to right) respectively.}}
	\label{rd_plot}
\end{figure*}
%%%%%%%%%%%%%%%%%%%%%

The statistical uncertainty $\mathbf{C}_{\mbox{\tiny stat}}$ and systematic uncertainty $\mathbf{C}_{\mbox{\tiny sys}}$ are also available. So, the total uncertainty matrix of distance modulus is, 
\begin{equation}
\mathbf{\Sigma}_\mu = \mathbf{C}_{\mbox{\tiny stat}} + \mathbf{C}_{\mbox{\tiny sys}}.
\end{equation}
The uncertainty of $d_L(z)$ is propagated from the uncertainties of $\mu$ using the standard error propagation formula,
\begin{equation}
\mathbf{\Sigma}_{d_L} = \mathbf{D}_1 \mathbf{\Sigma}_\mu {\mathbf{D}_1}^{\mbox{\small T}} 
\end{equation}
where the superscript `${\mbox{\small T}}$' denotes the transpose of a matrix. Also, $\mathbf{D}_1$ is a Jacobian matrix given by 
\begin{equation}
\mathbf{D}_1 = \mbox{diag}\left(\frac{\ln 10}{5} \mathbf{d_L}\right),
\end{equation}
where $\mathbf{d_L}$ is a vector whose components are the luminosity distances of all SN-Ia.\\

We know the absolute magnitude of SN-Ia is degenerate with the Hubble parameter, so we obtain the marginalized constraints on $M_B$ following a similar procedure as in \citet{purba_jerk}.

\subsection{$D_V(z)$ from BAO data}

The baryon acoustic oscillations are regular, periodic fluctuations in the visible baryonic matter density, and are widely used as ``standard rulers" to measure the distances in cosmology. Here we use data points from the following surveys:
\begin{enumerate}
	\item 6dFGS \citep{beutler2011} at effective redshift $z=0.106$.
	\item WiggleZ Dark Energy Survey \citep{blake2012} at effective redshifts $z = 0.44$, $0.6$ and $0.73$.
	\item MGS and SDSS (LOWZ and CMASS samples) \citep{xu2013, anderson2014, ross2015, gilmarin2015} at effective redshift $z = 0.15, 0.32, 0.57$.
	\item BOSS DR12 \citep{alam2017} at $z = 0.38, 0.51, 0.61$.
	\item eBOSS DR14 LRG \citep{bautista2018} and quasar \citep{ata2018} sample at redshifts $z = 0.72, 1.52$ respectively.
%	\item DES \citep{abbott2019} at $z = 0.81$.
	\item HIGHZ Lyman-$\alpha$ measurement in BOSS DR12 \citep{bautista2017} at redshift $z = 2.33$.
%	\item Quasar-Lyman-$\alpha$ Forest Cross-Correlation from BOSS DR11 at $z = 2.36$ \citep{font-ribera}.
	\item Correlations of Ly$\alpha$ absorption in eBOSS DR14 at $z=2.34$ \citep{agathe}.
	\item Cross-correlation of Ly$\alpha$ absorption and quasars in eBOSS DR14 at $z=2.35$  \citep{blomqvist}.
\end{enumerate}
For the purpose of our analysis, we shall use the volume-averaged $D_V$ distance data. The WriggleZ DES data is cosmological parameter dependent ($\Omega_{m0}$ and $H_0$), hence the uncertainties associated is propagated in the evaluation of the distance errors, that are used for the GP evaluation. We define, \begin{equation} \label{Dv}
	D_V (z)= \left[(1+z)^2 d_A^2(z) \frac{c z}{H(z)}\right]^\frac{1}{3},
\end{equation} therefore \begin{equation}
	d_A(z) = \left[\frac{D_V^3(z)~ H(z)}{c ~z~ (1+z)^2}\right]^\frac{1}{2}
\end{equation} and $r_{\mbox{\tiny d, fid}} = 147.49$ is considered unless explicitly specified in the datasets. The comoving sound horizon at the drag epoch $r_d$ is defined as \citep{rdrag_def}, \begin{equation} \label{rdrag_def}
	r_d = \int_{z_{\mbox{\tiny d}}}^{\infty} \frac{c_s(z)}{H(z)} dz
\end{equation} with $c_s(z)$ the sound speed and $z_{\mbox{\tiny d}}$ the redshift at the drag epoch. Equation \eqref{rdrag_def} can be approximated as, \begin{equation} \label{rdrag_approx}
	r_d  \simeq  \frac{44.5 \log \left(\frac{9.83}{\Omega_{m0} h_0^2}\right)}{\sqrt{1+10\left(\Omega_{b0}h_0^2\right)^{3/4}}}
\end{equation} for the standard model cosmology where we rewrite, $h_0 = \frac{H_0}{100}$ in a dimensionless way. In this particular work, constraints on the comoving sound horizon at the drag epoch $r_d$ (in Mpc) has been obtained keeping it a free parameter in our analysis.

%%%%%%%%%%%%%%%%%%%%%
\begin{table*} 
	\caption{{\small Table showing the value of $r_d$ (in units of Mpc) for different choices of the covariance function, considering the approximated definition from equation \eqref{rdrag_approx}.}}
	\begin{center}
		\resizebox{0.98\textwidth}{!}{\renewcommand{\arraystretch}{1.3} \setlength{\tabcolsep}{20pt} \centering  
			\begin{tabular}{c c c c c} 
				%				\hline
				\hline
				$k(z,\tilde{z})$ & Mat\'{e}rn 9/2 & Mat\'{e}rn 7/2 & Mat\'{e}rn 5/2 &  Squared Exponential\\ 
				\hline
				\hline
				$r_d$  &  $149.828 \pm 12.059$  & $149.959 \pm 12.313$ & $150.134 \pm 12.921$	& $149.181 \pm 11.496$\\ 
				\hline
			\end{tabular}
		}
	\end{center}
	\label{rdrag_tab}
\end{table*}

\begin{figure*}%
	\begin{center}
		\includegraphics[angle=0, width=0.245\textwidth]{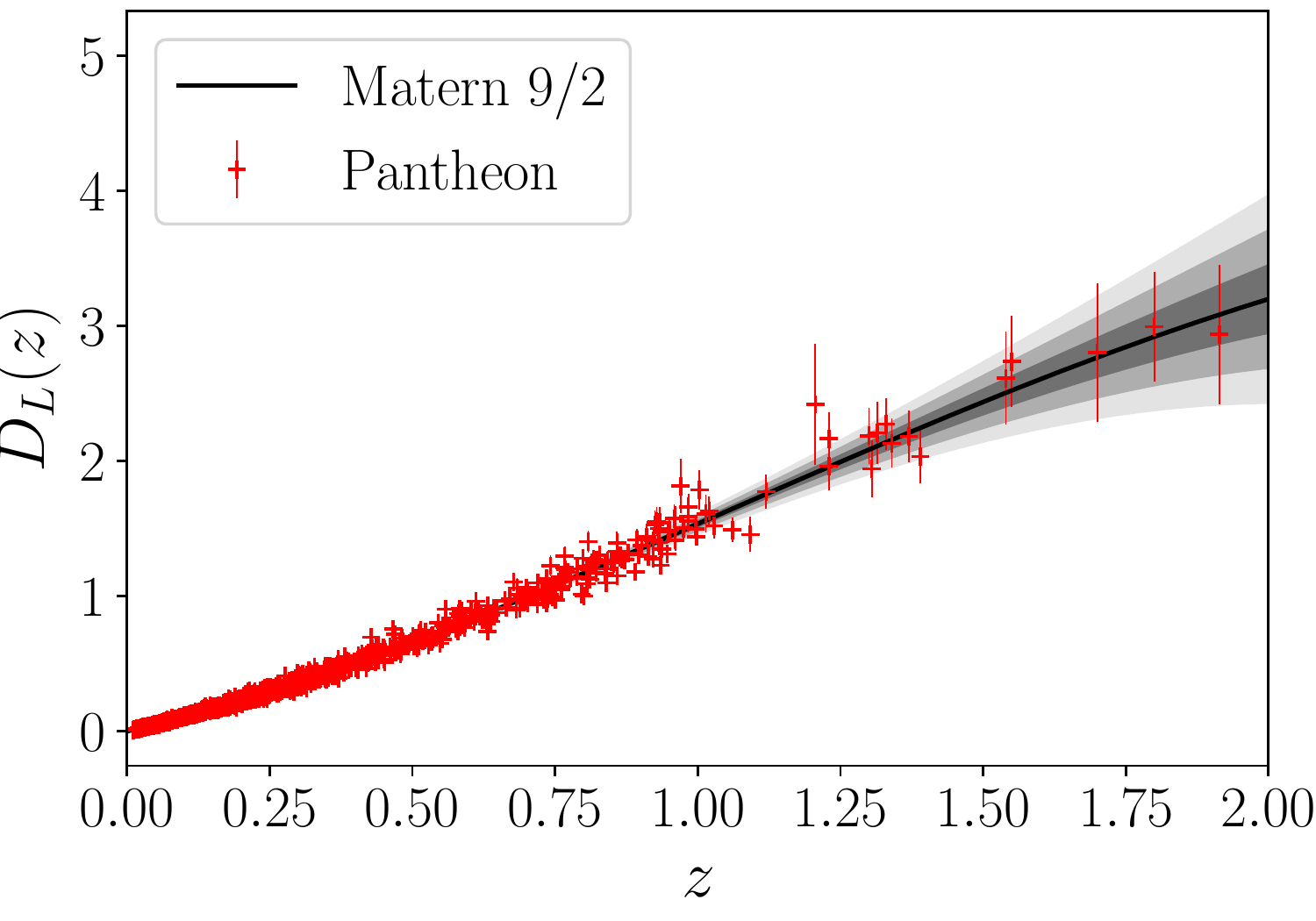}
		\includegraphics[angle=0, width=0.245\textwidth]{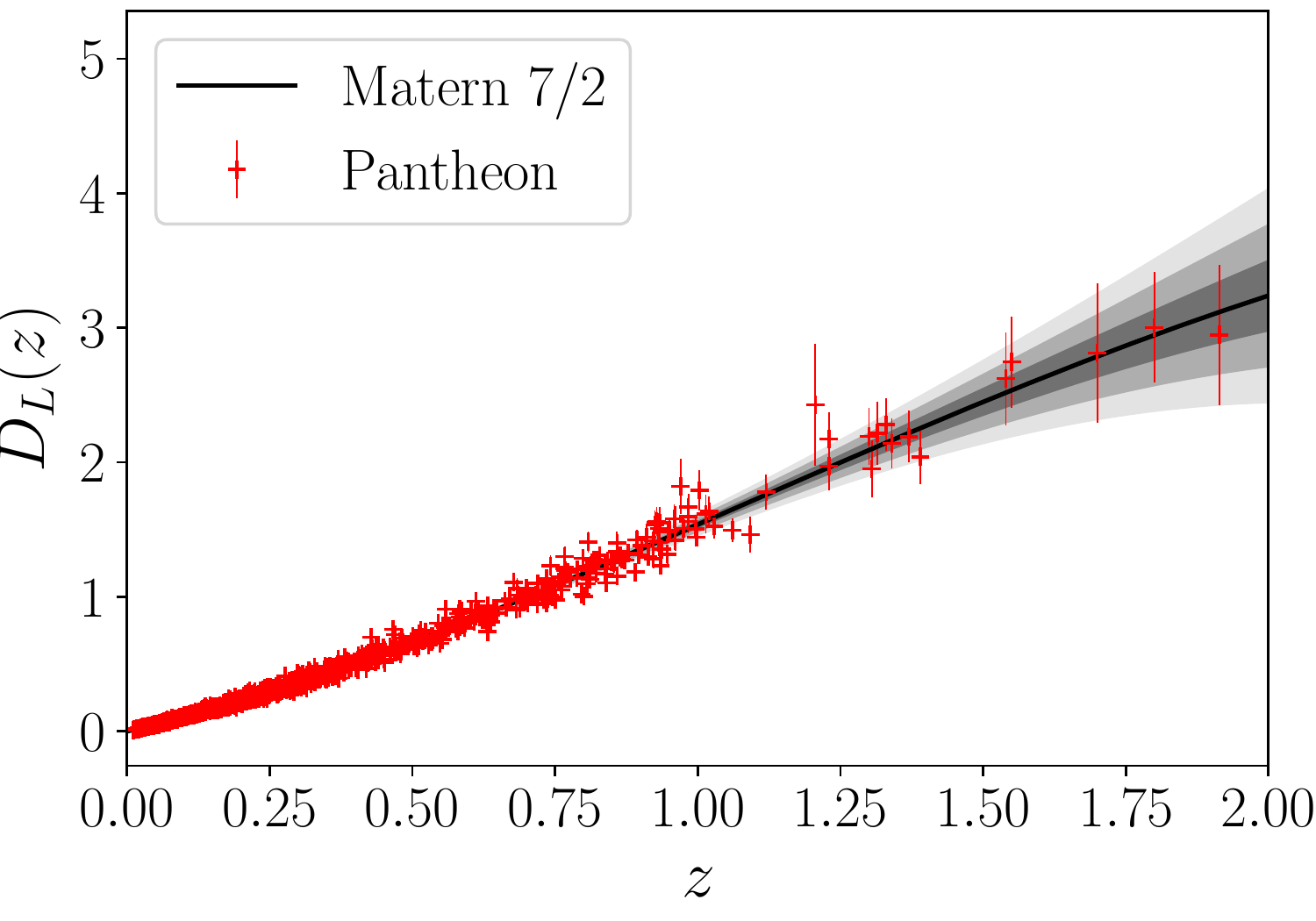}
		\includegraphics[angle=0, width=0.245\textwidth]{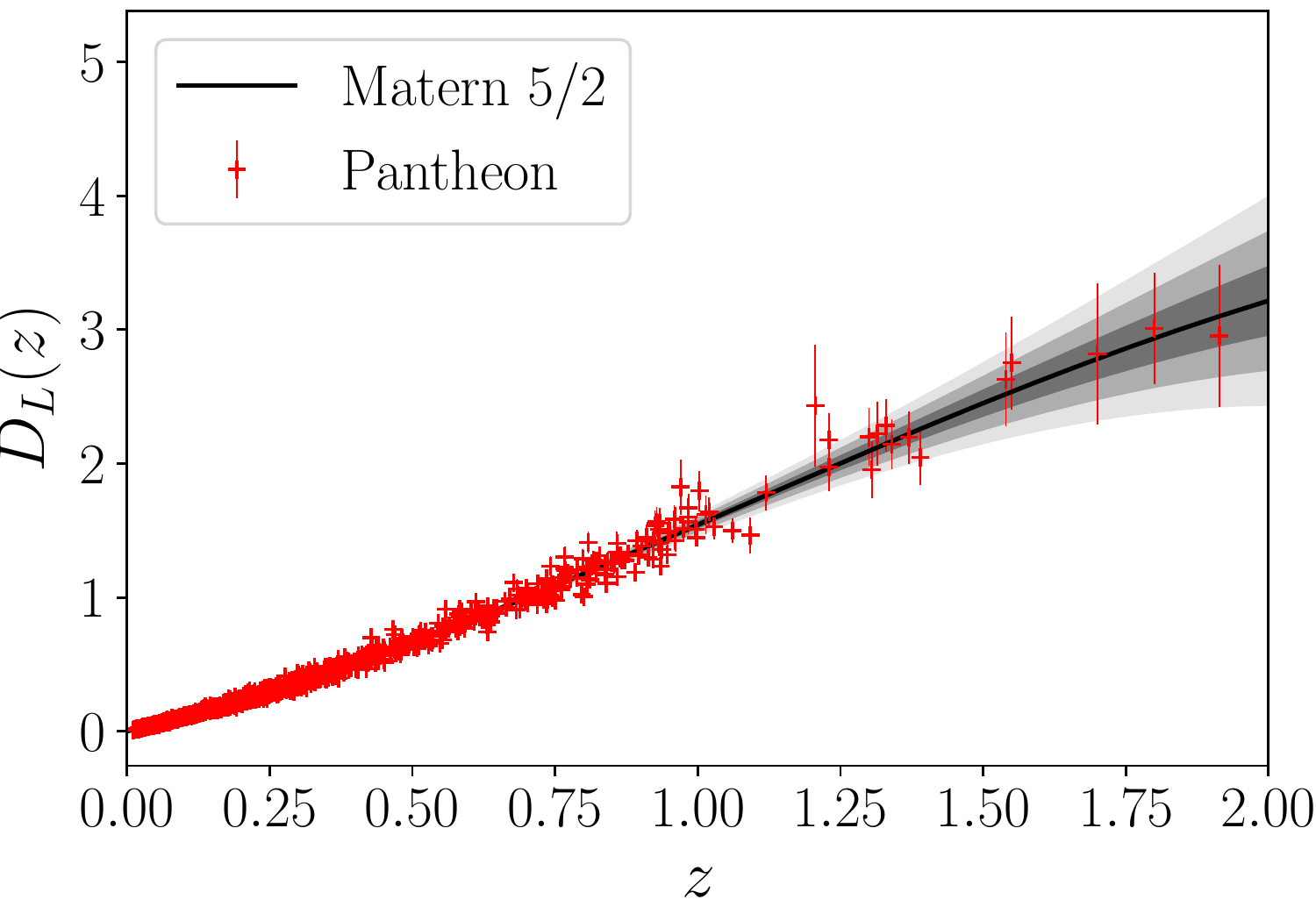}
		\includegraphics[angle=0, width=0.245\textwidth]{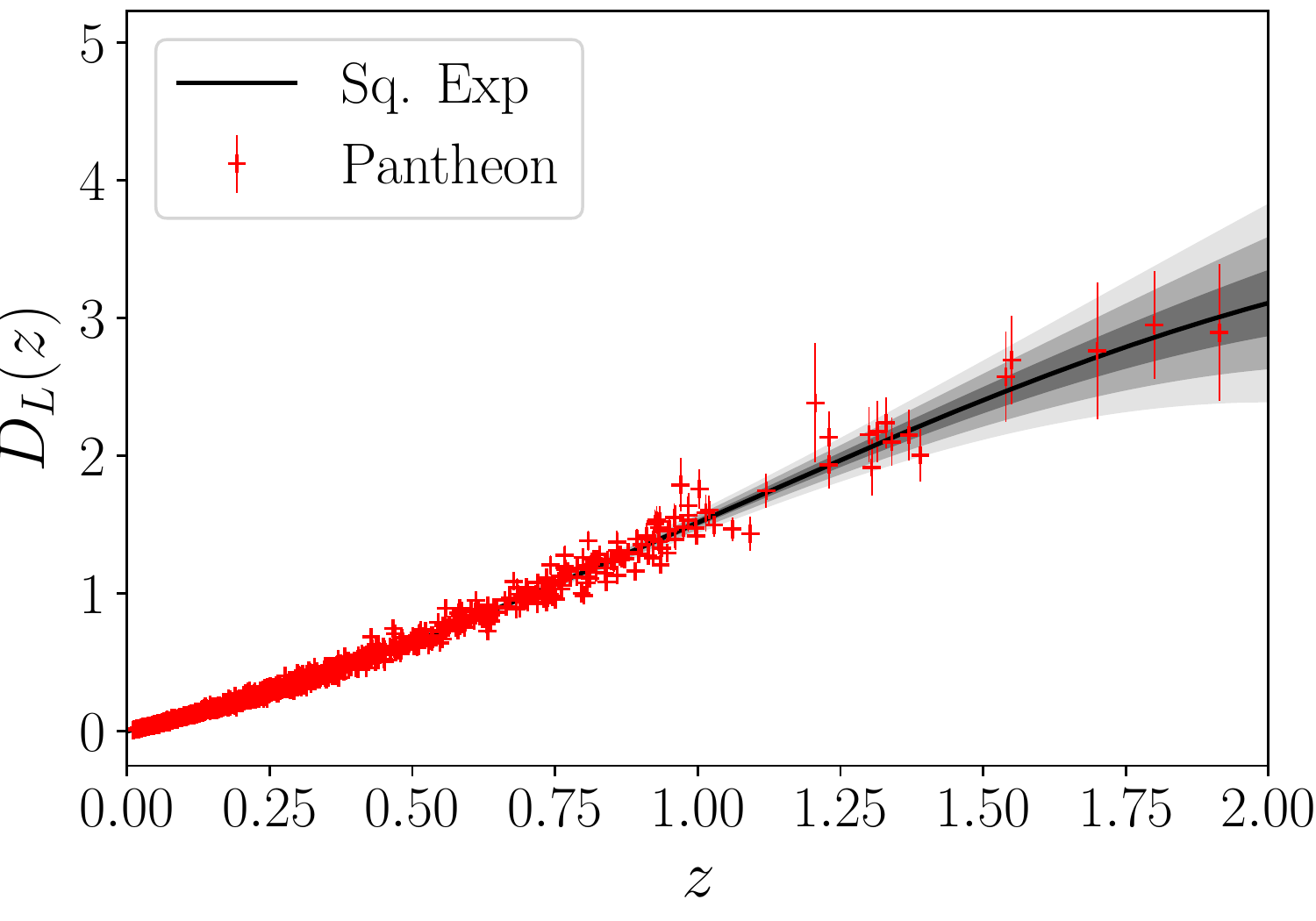}
	\end{center}
	\caption{{\small Plots for the reconstructed dimensionless or normalized luminosity distance $D_L$ considering the Mat\'{e}rn 9/2, Mat\'{e}rn 7/2, Mat\'{e}rn 5/2 and Squared Exponential (from left to right) covariance function. The black solid lines represent the best fitting curves. The associated 1$\sigma$, 2$\sigma$ and 3$\sigma$ confidence levels are shown by the shaded regions.}}
	\label{Dl_plot}
\end{figure*}
%%%%%%%%%%%%%%%%%%%%%

%%%%%%%%%%%%%%%%%%%%%
\begin{figure*}%
	\begin{center}
		\includegraphics[angle=0, width=0.245\textwidth]{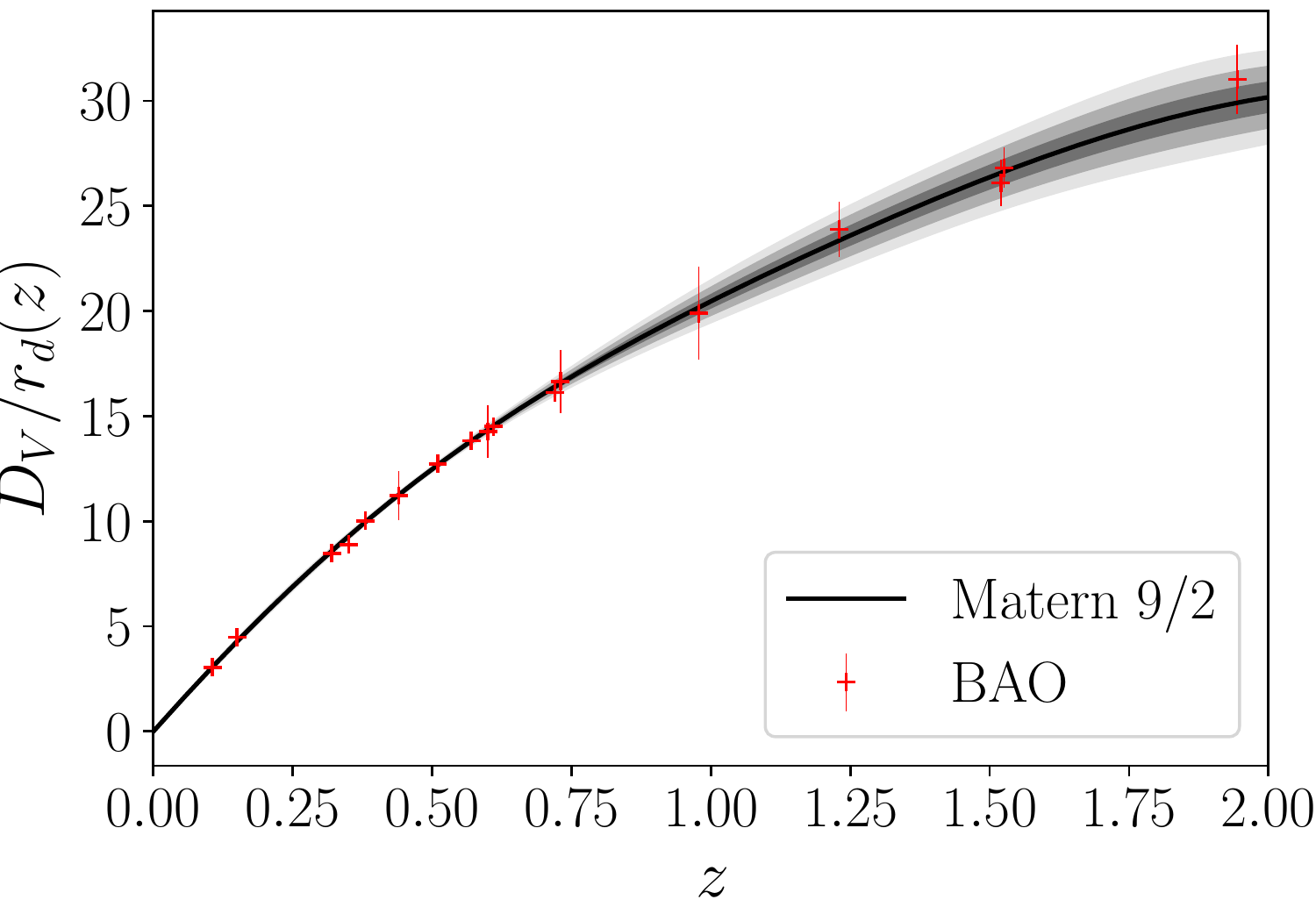}
		\includegraphics[angle=0, width=0.245\textwidth]{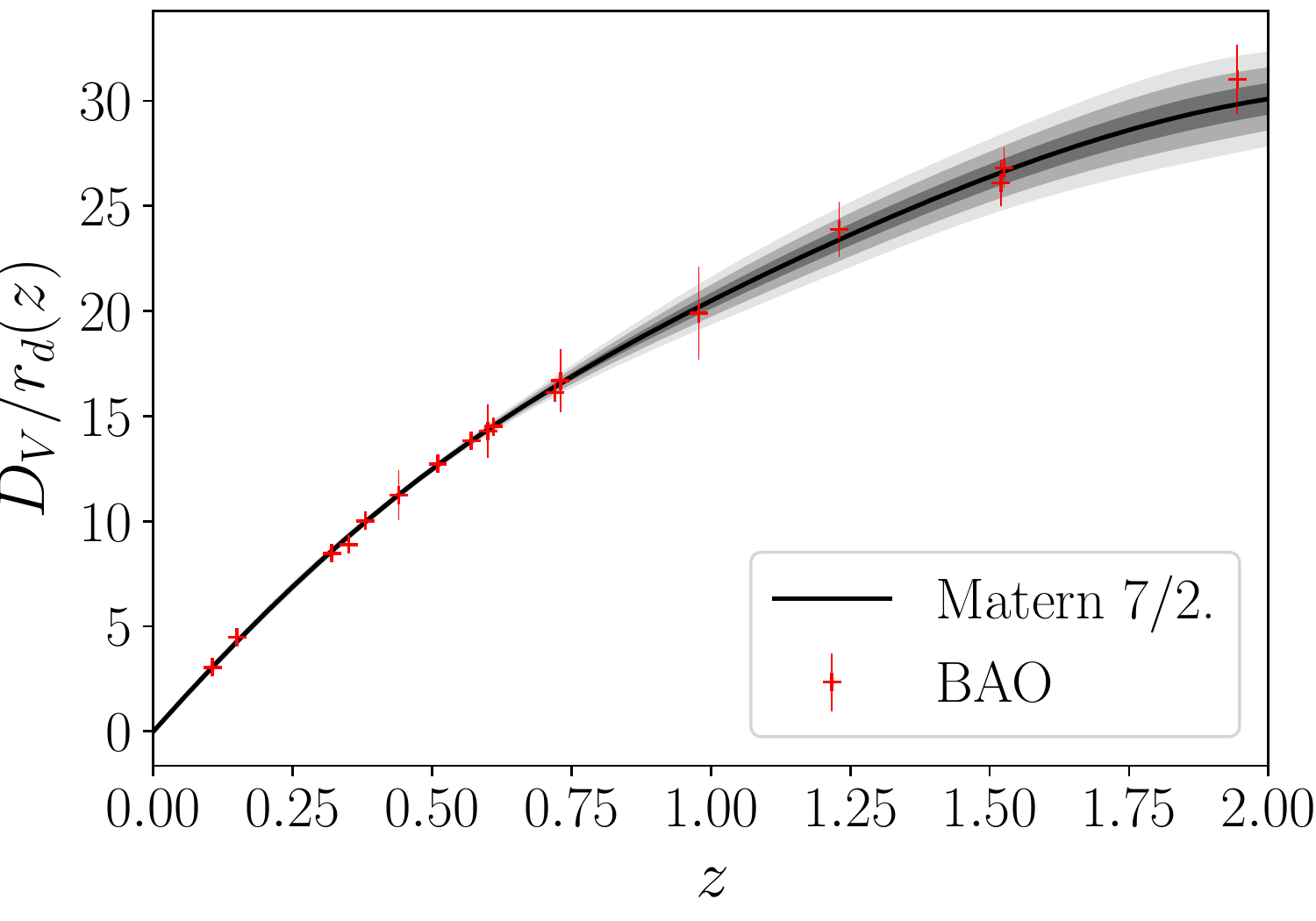}
		\includegraphics[angle=0, width=0.245\textwidth]{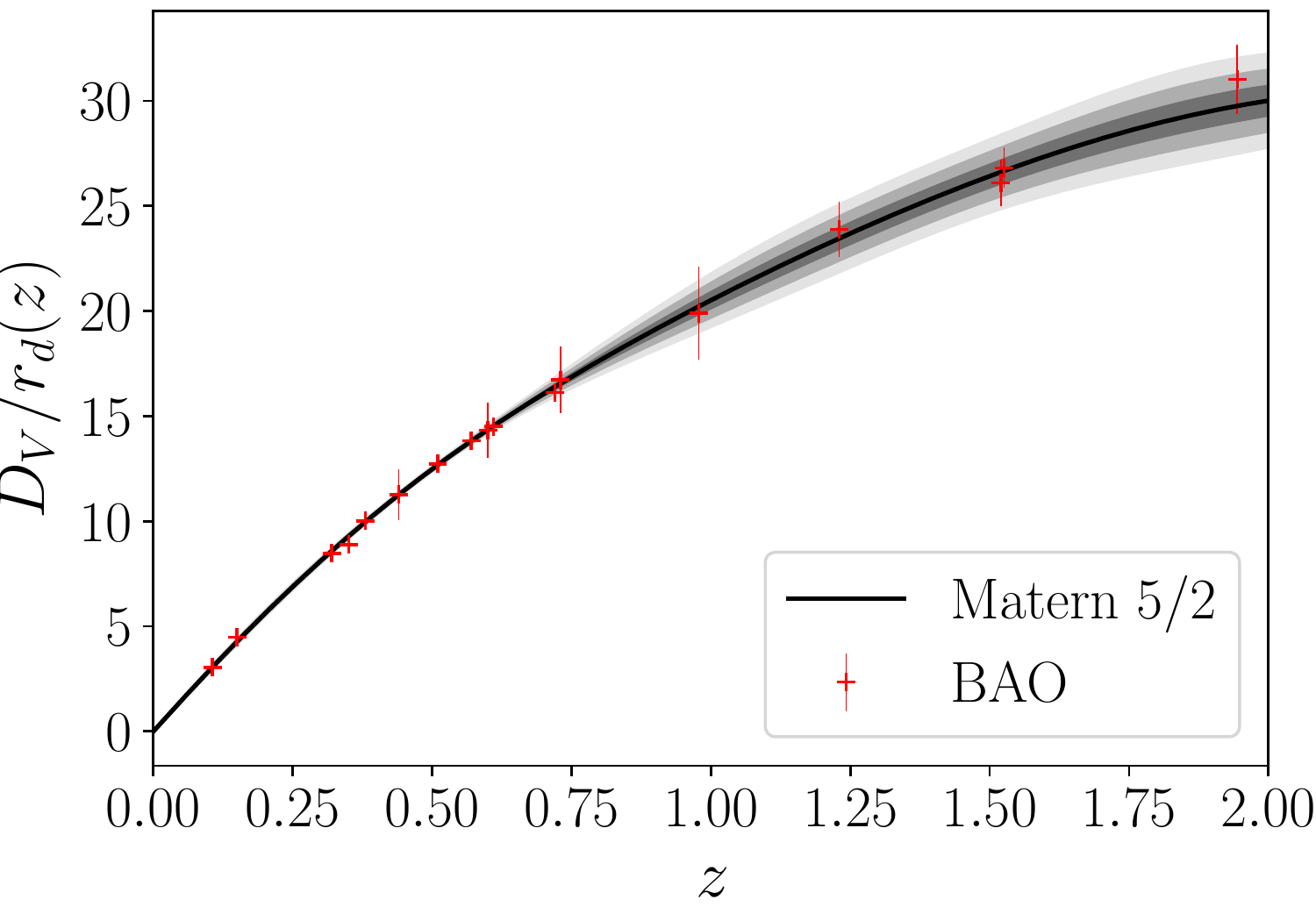}
		\includegraphics[angle=0, width=0.245\textwidth]{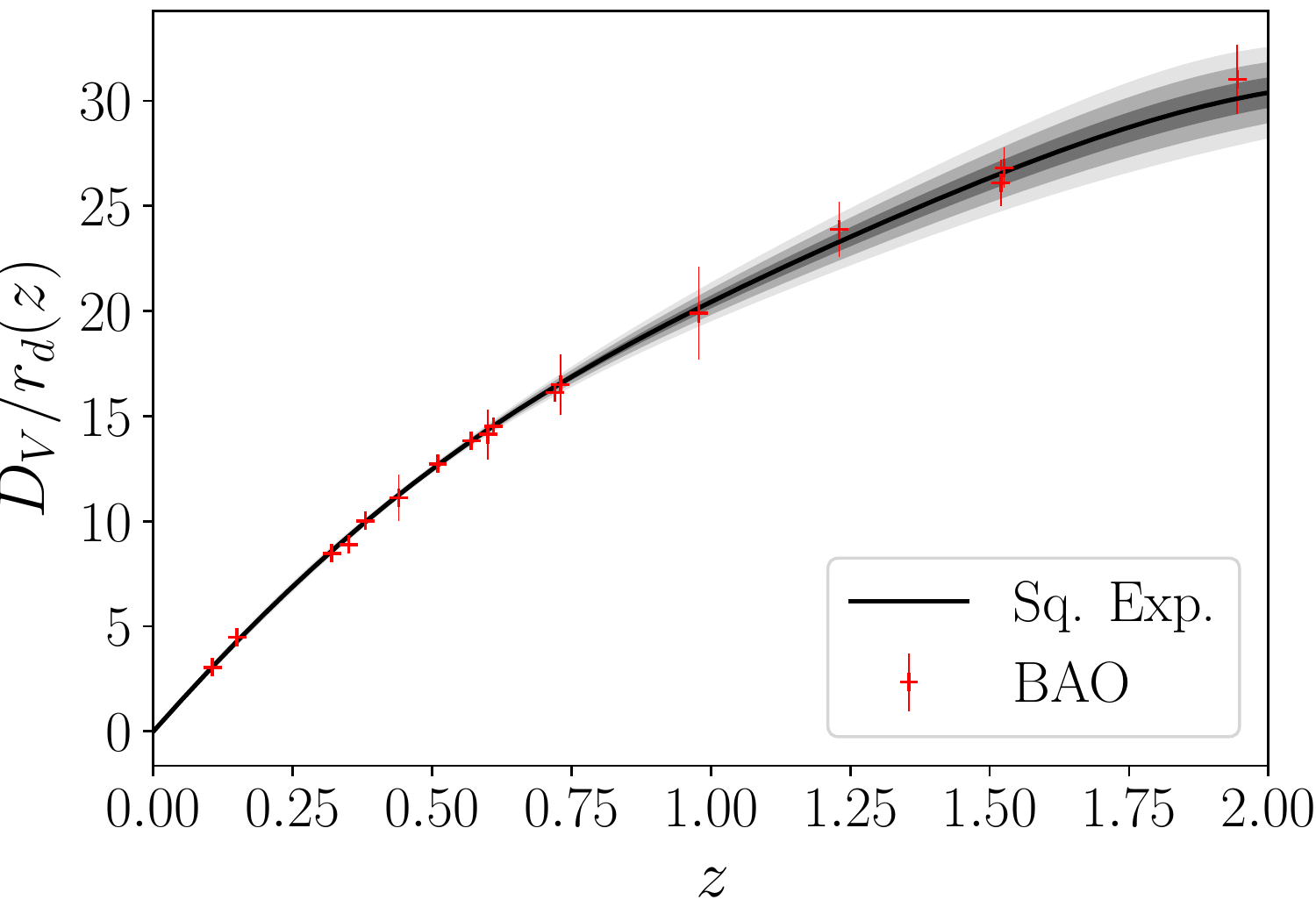}
	\end{center}
	\caption{{\small Plots for the reconstructed dimensionless ratio of volume-averaged distance to the comoving sound horizon at drag epoch $\frac{D_V}{r_d}$ considering the Mat\'{e}rn 9/2, Mat\'{e}rn 7/2, Mat\'{e}rn 5/2 and Squared Exponential (from left to right) covariance function. The black solid lines represent the best fitting curves. The associated 1$\sigma$, 2$\sigma$ and 3$\sigma$ confidence levels are shown by the shaded regions.}}
	\label{Dv_plot}
\end{figure*}

\subsection{$H(z)$ from CC data}

The Hubble parameter $H(z)$ can be directly obtained from the differential redshift time derivative, by calculating the spectroscopic differential ages of passively evolving galaxies, usually called the Cosmic Chronometer (CC) method. In this work we use the 32 CCH data points measured by considering the BC03 stellar population synthesis model \citep{cc0, cc2, cc3, cc1, cc4, cc5, cc6}, covering the redshift range up to $z \sim 2$. These measurements do not assume on any particular cosmological model.

\section{Results}
\label{results}

%%%%%%%%%%%%%%%%%%%%%
\begin{figure*}%
	\begin{center}
		\includegraphics[angle=0, width=0.245\textwidth]{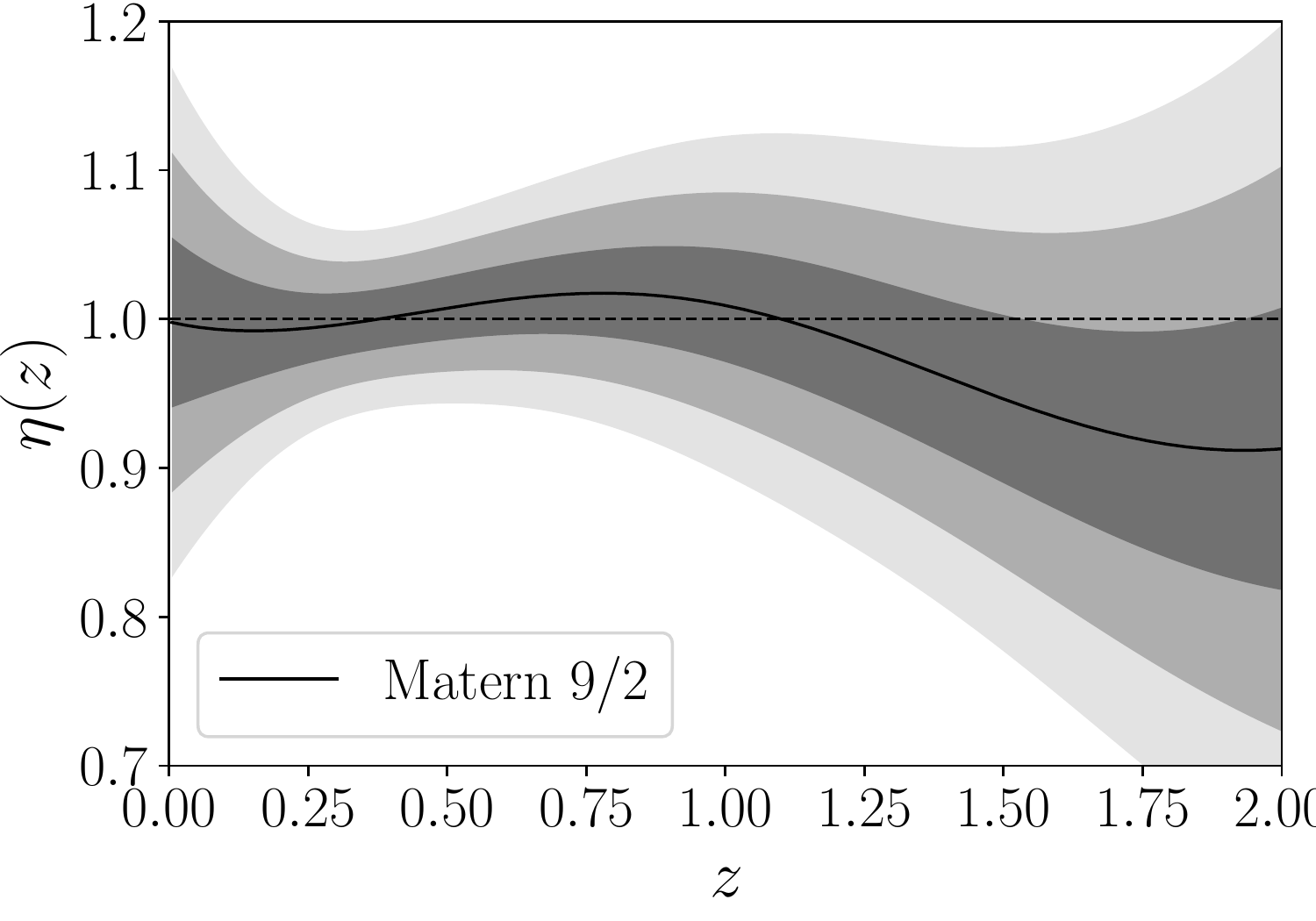}
		\includegraphics[angle=0, width=0.245\textwidth]{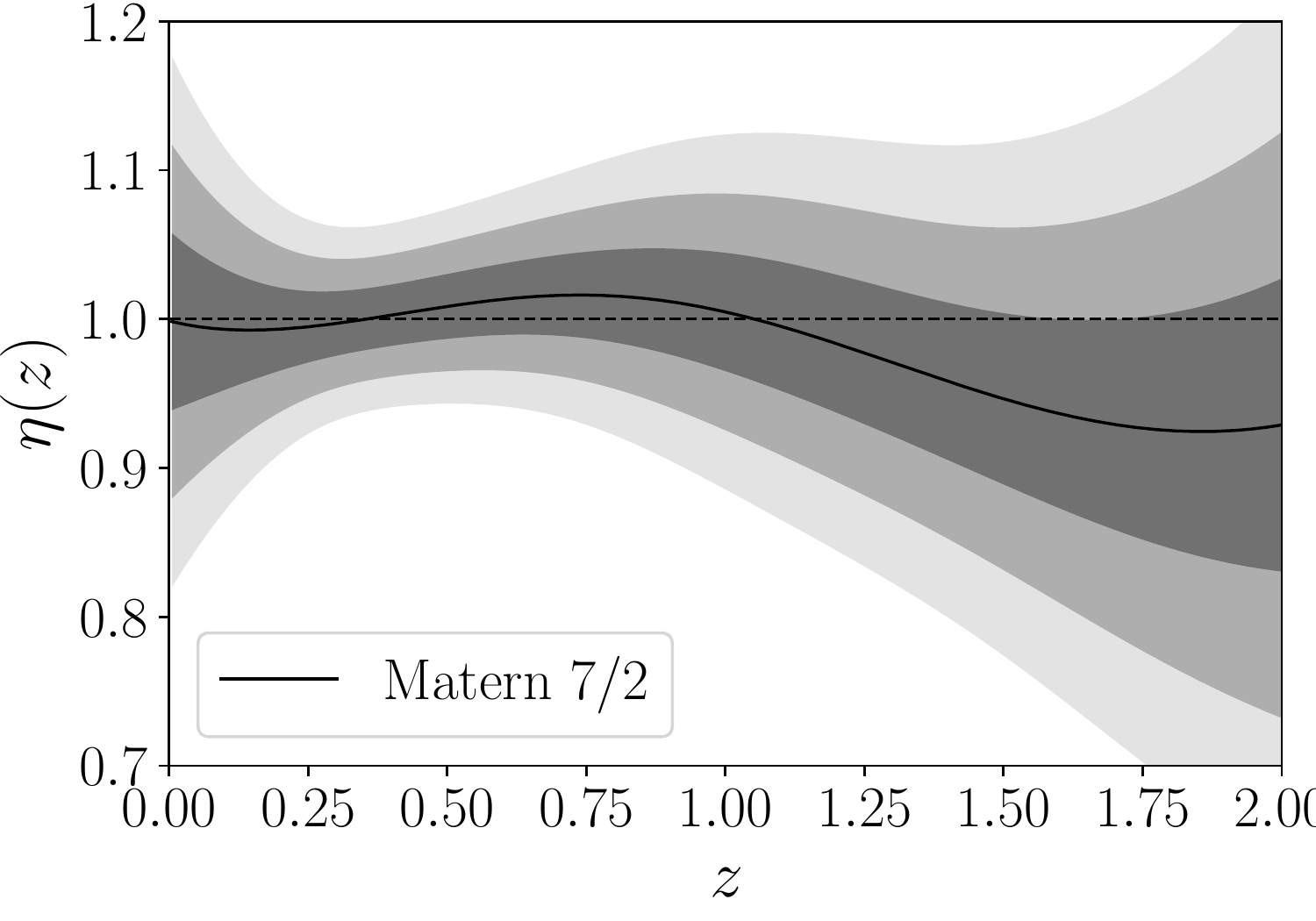}
		\includegraphics[angle=0, width=0.245\textwidth]{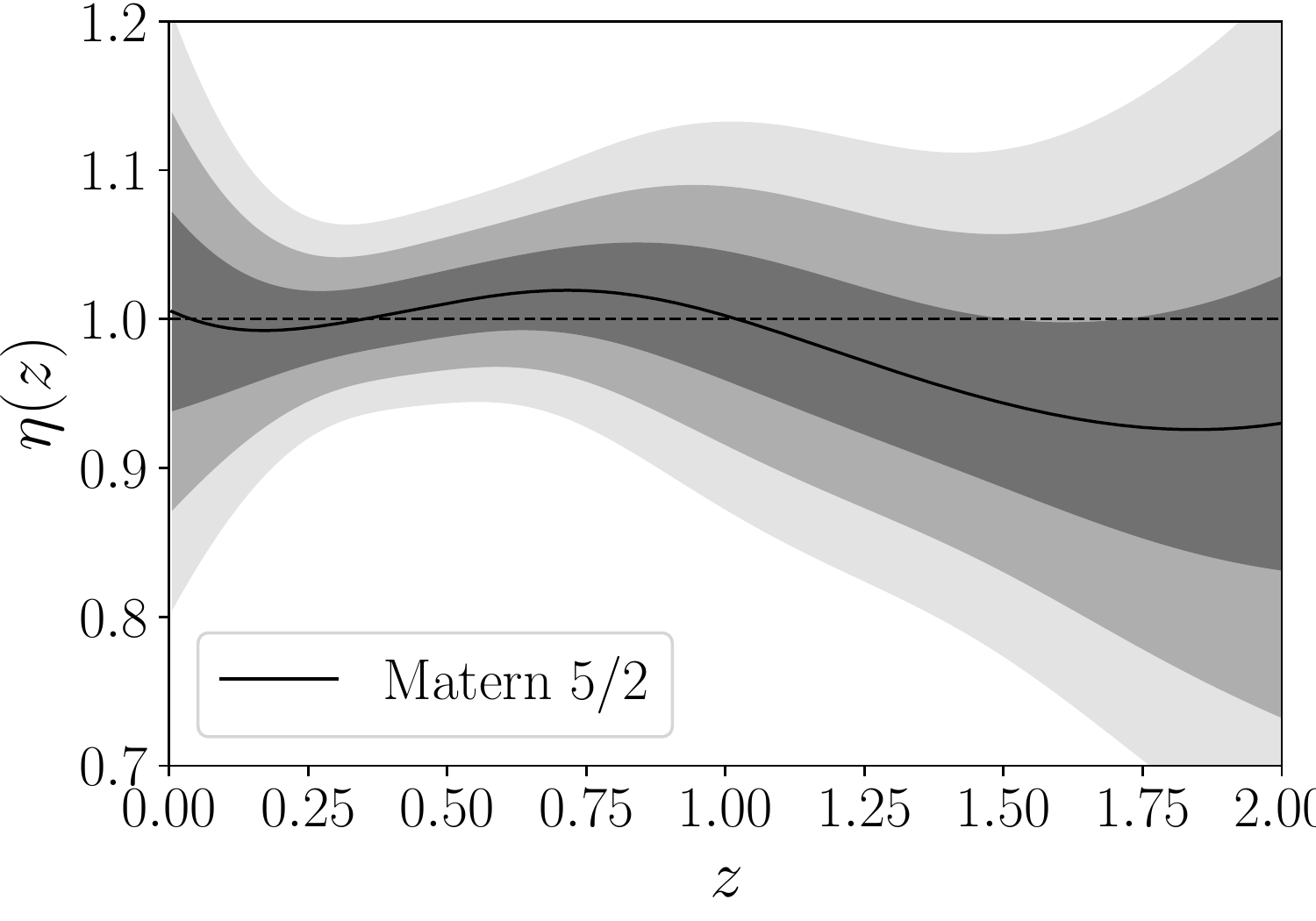}
		\includegraphics[angle=0, width=0.245\textwidth]{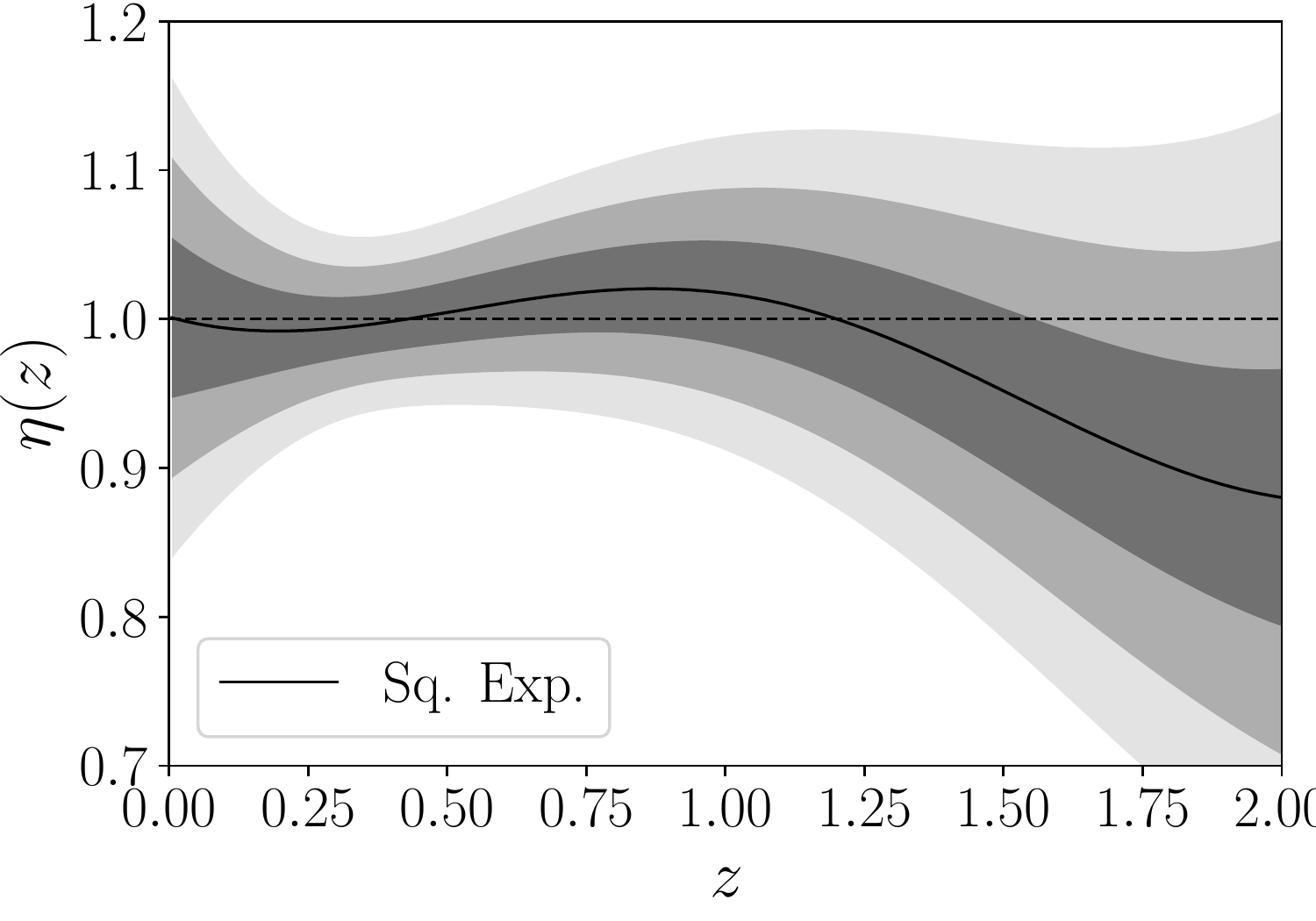}
	\end{center}
	\caption{{\small Plots for the reconstructed cosmic distance duality ratio $\eta$ considering the Mat\'{e}rn 9/2, Mat\'{e}rn 7/2, Mat\'{e}rn 5/2 and Squared Exponential (from left to right) covariance function using $r_d$ from Table \ref{MOr_res}. The black solid lines represent the best fitting curves. The associated 1$\sigma$, 2$\sigma$ and 3$\sigma$ confidence levels are shown by the shaded regions.}}
	\label{eta_np_plot}
\end{figure*}

%%%%%%%%%%%%%%%%%%%%%
\begin{figure*}%
	\begin{center}
		\includegraphics[angle=0, width=0.245\textwidth]{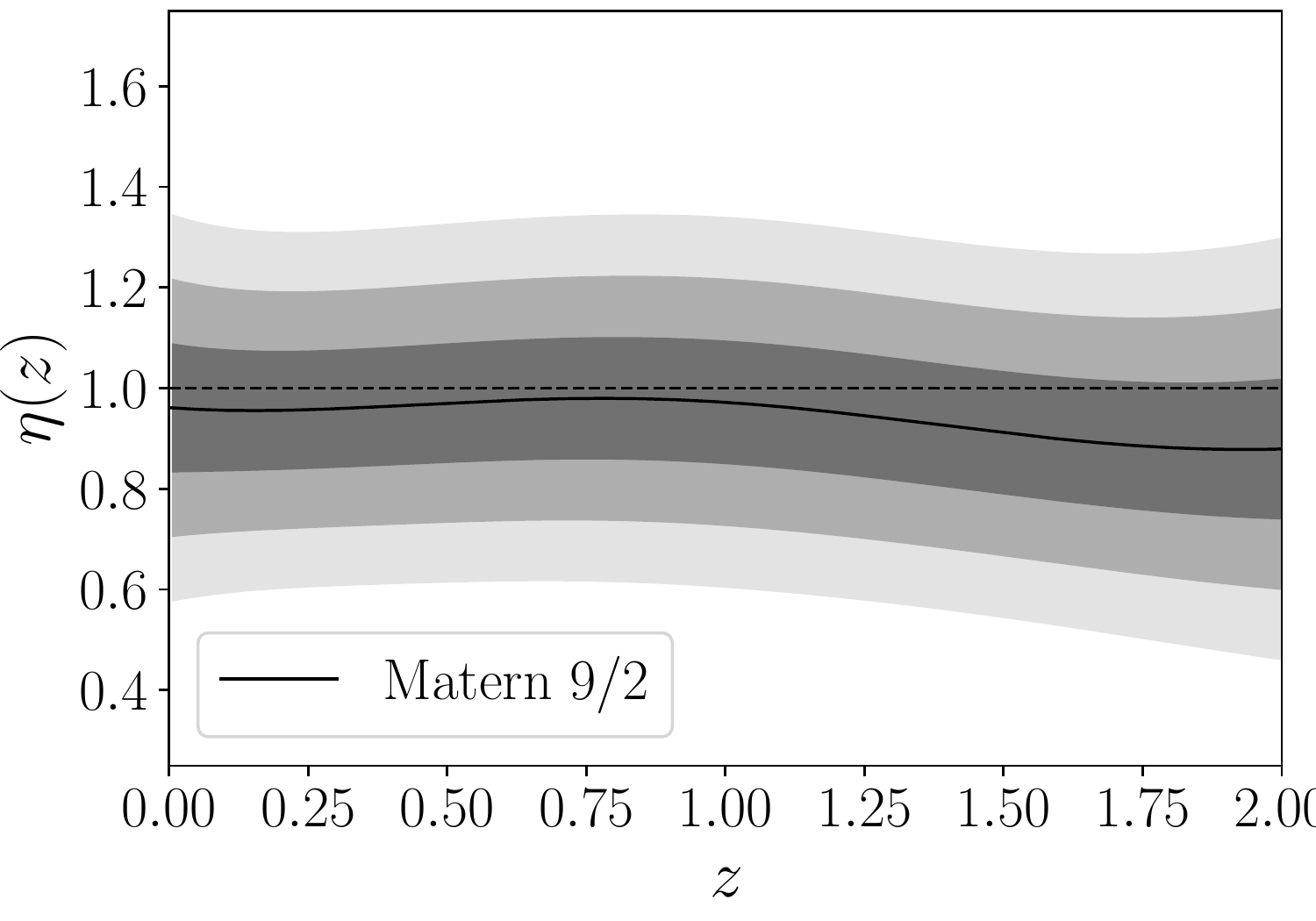}
		\includegraphics[angle=0, width=0.245\textwidth]{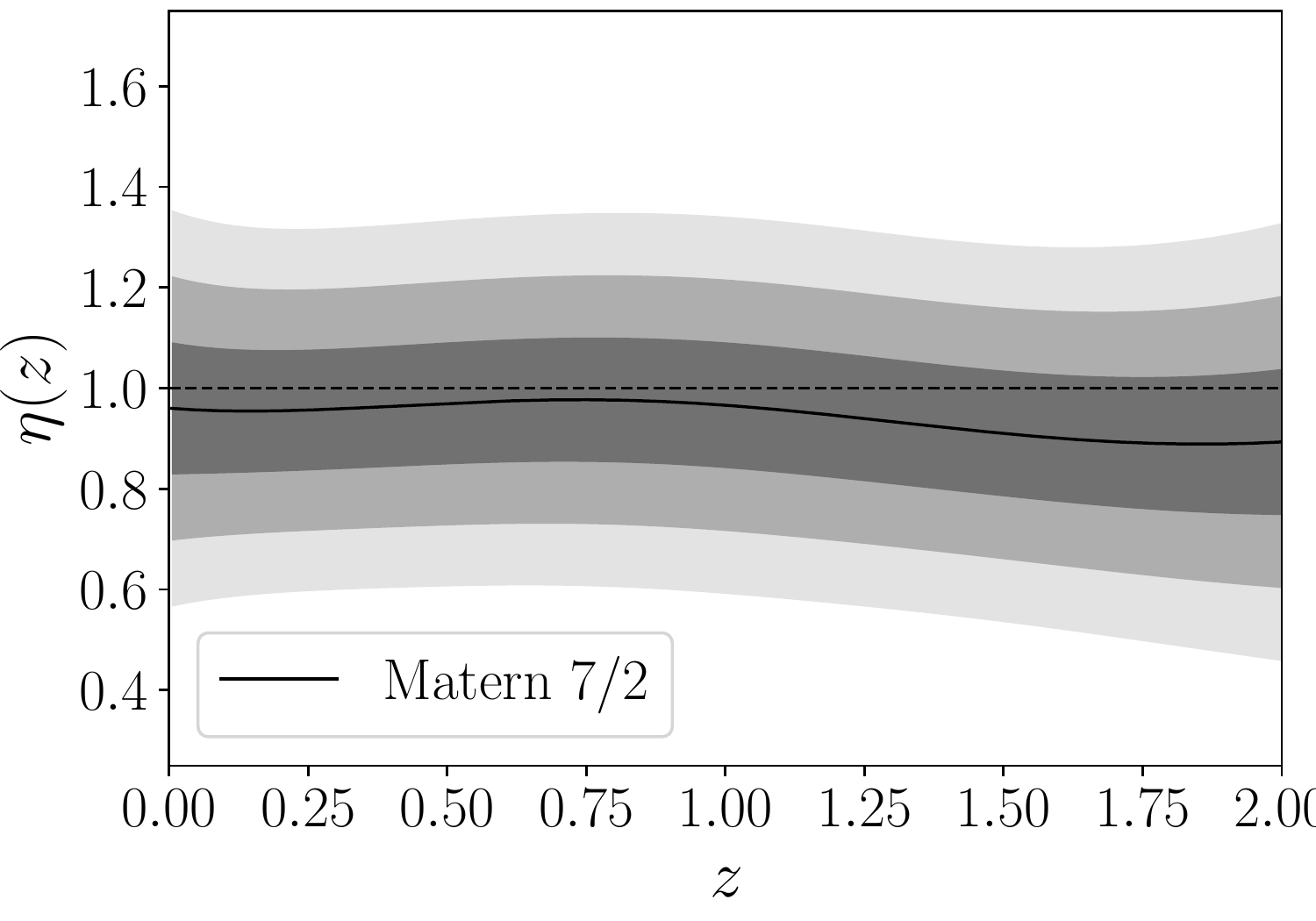}
		\includegraphics[angle=0, width=0.245\textwidth]{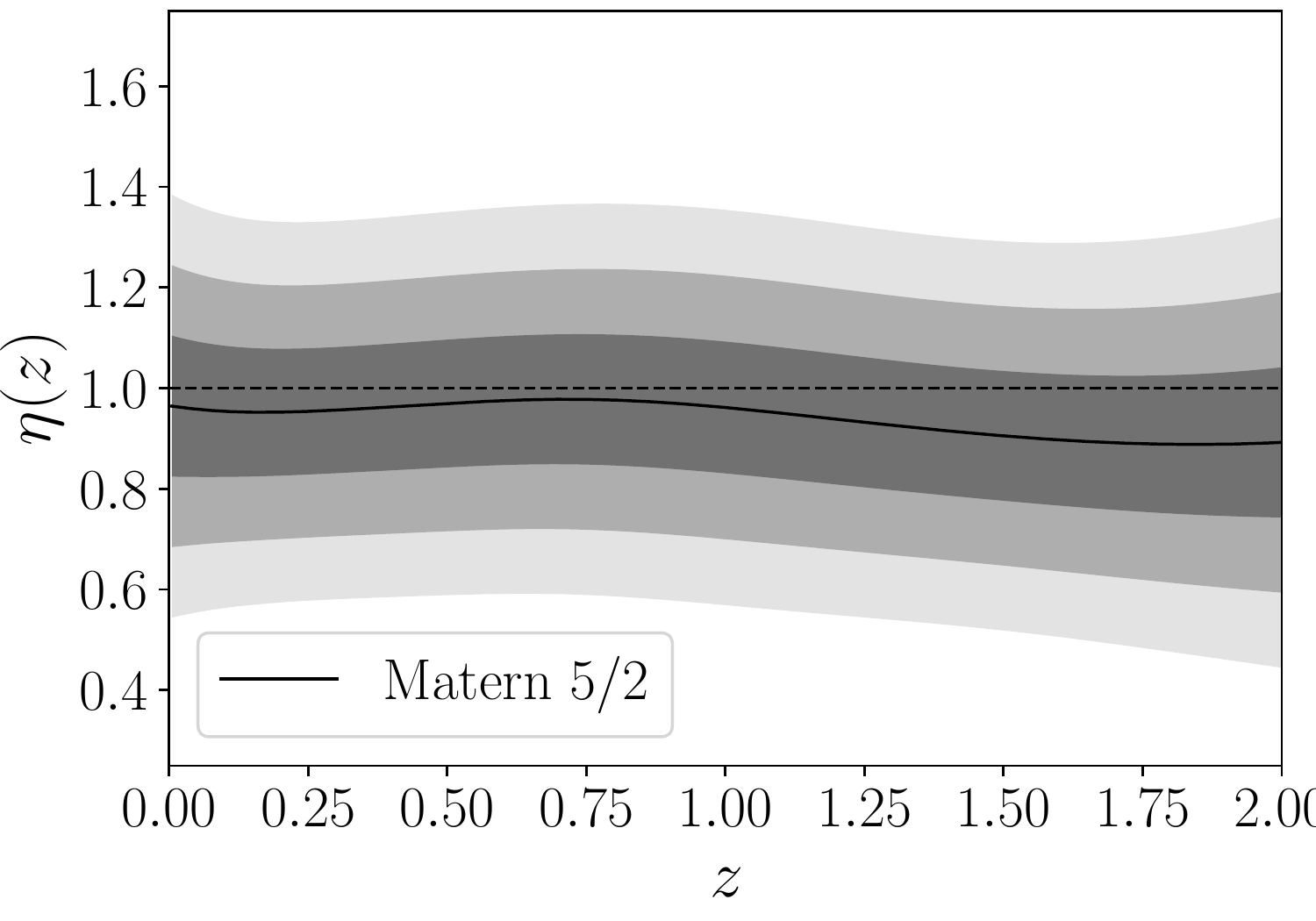}
		\includegraphics[angle=0, width=0.245\textwidth]{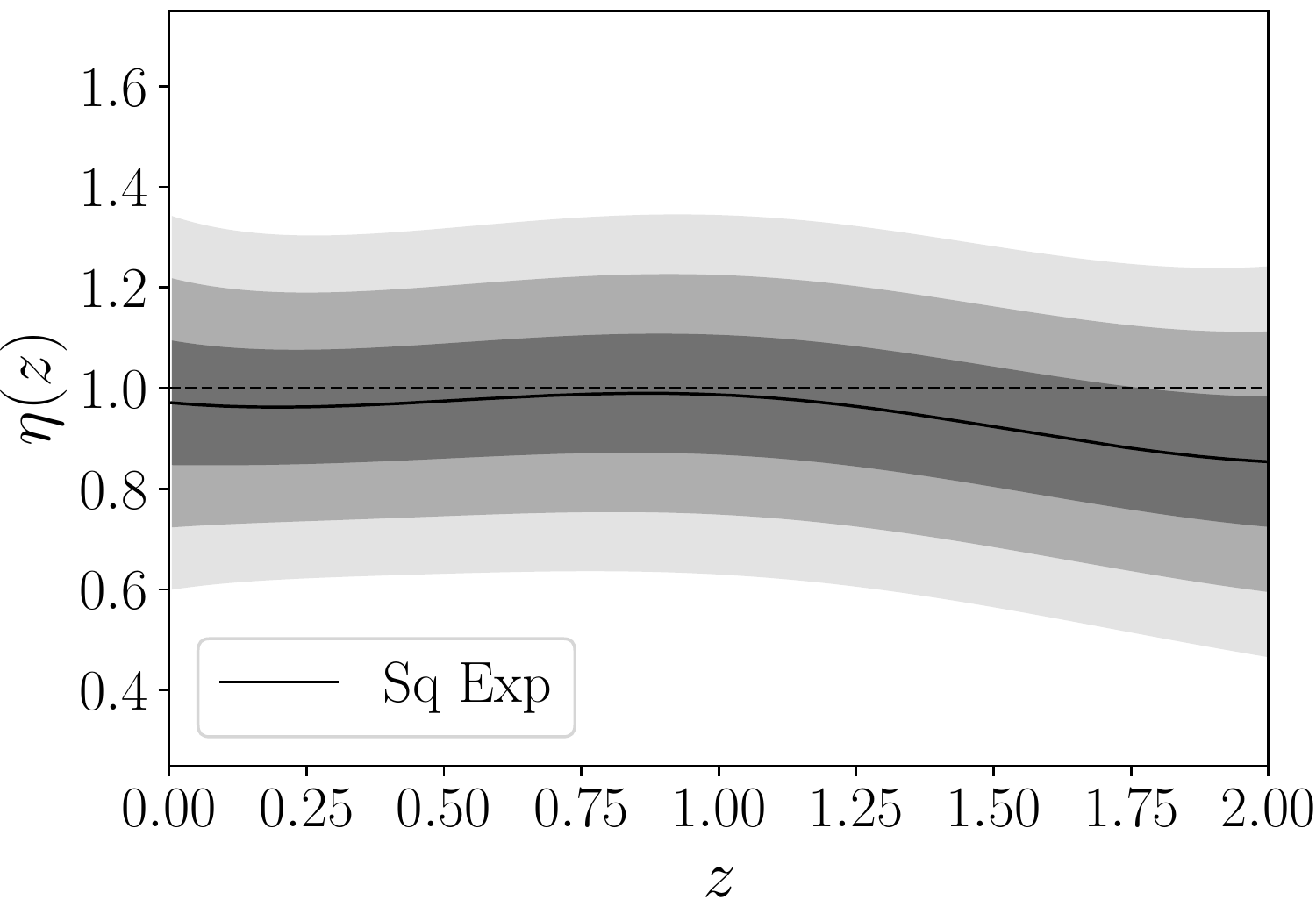}
	\end{center}
	\caption{{\small Plots for the reconstructed cosmic distance duality ratio $\eta$ considering the Mat\'{e}rn 9/2, Mat\'{e}rn 7/2, Mat\'{e}rn 5/2 and Squared Exponential (from left to right) covariance function using $r_d$ from Table \ref{rdrag_tab}. The black solid lines represent the best fitting curves. The associated 1$\sigma$, 2$\sigma$ and 3$\sigma$ confidence levels are shown by the shaded regions.}}
	\label{eta_np_plot2}
\end{figure*}
%%%%%%%%%%%%%%%%%%%%%

We reconstruct the Hubble parameter $H(z)$ using the GP method and the results are shown Fig. \ref{Hz_recon}. The value of the Hubble constant $H_0$ obtained from this model independent reconstruction is shown in Table \ref{Hz_res}. Again, we normalize the reconstructed $H(z)$ data to obtain the dimensionless or reduced Hubble parameter $E(z) = H(z)/H_0$. We calculate the uncertainty in $E(z)$ by the standard technique of error propagation as,
\begin{equation} \label{sig_h}
{\sigma_{E}}^2 = \frac{{\sigma_H}^2}{ {H_0}^2} + \frac{H^2}{{H_0}^4}{\sigma_{H_0}}^2,
\end{equation} 
where $\sigma_{H_0}$ is the error associated with $H_0$.\\

Utilizing the reconstructed $E(z)$ function, the normalised comoving distance $D$ for a flat spacetime is evaluated numerically using the trapezoidal integration rule such that,
\begin{equation} \label{D_recon}
D(z) = \int_{0}^{z} \frac{d z'}{E(z')}.
\end{equation} The error associated with $D$, $\sigma_{D}$ is obtained by following a similar integration of the function $E(z)$ including it's 1$\sigma$ error uncertainties.\\

We perform another GP reconstruction on the apparent magnitudes $m_B$ of the SN-Ia data and reconstruct them at the same redshift $z$ as that of the CC data. On substituting equation \eqref{D_recon} in equation \eqref{mu}, we estimate the reconstructed distance modulus from the Hubble data as, $\mu_{\mbox{\tiny H}} = 5 \log_{10} \left[D(1+z)\right] + 25$ along with it's 1$\sigma$ error uncertainty $\sigma_{\mu_{\mbox{\tiny H}}}$ given by,
\begin{equation}
\sigma_{\mu_{\mbox{\tiny H}}} = \frac{5}{\ln 10} \frac{\sigma_{D}}{D}.
\end{equation}

The absolute magnitude of SN-Ia is degenerate with the Hubble parameter $H_0$. We get the marginalized constraints on $M_B$ by minimizing the $\chi^2$ function,  \begin{equation} \label{chiSN}
\chi_{\mbox{\small SN}}^2 = \sum \Delta \mu^{\mbox{\small T}} \cdot \Sigma^{-1} \cdot \Delta \mu,
\end{equation} 
considering a uniform prior $M_B$ $\in [-35,-5]$, where $\Delta \mu = (\mu_{\mbox{\tiny SN }}- \mu_{\mbox{\tiny H}})$ and $\Sigma = \Sigma_{\mu_{\mbox{\tiny SN }}} + \sigma_{\mu_{\mbox{\tiny H}}}^2$ respectively.\\

Further we perform the reconstruction of $\frac{D_V}{r_d}$ from the BAO data compilation in the same redshift range $0<z<2$. The WriggleZ DES data measures the acoustic scale parameter $A(z)$ given by,
\begin{equation}
	A(z) = D_V \frac{\sqrt{\Omega_{m0} ~ H_0^2}}{c z}
\end{equation} As the matter density parameter $\Omega_{m0}$ is correlated with $H_0$ we constrain $\Omega_{m0}$ assuming a fiducial $\Lambda$CDM model, \begin{equation}\label{chiCC}
	\chi_{\mbox{\small H}}^2 = \sum_i \frac{\left[ E(z_i) -  \sqrt{\Omega_{m0}(1+z_i)^3 + 1-\Omega_{m0}} \right]^2}{\sigma_{E}^2(z_i)}
\end{equation}
with uniform prior $\Omega_{m0} \in [0.01,0.7]$ using the reconstructed $E(z)$ data, for the same $H_0$ obtained in Table \ref{Hz_res}. For the remaining $BAO$ data sets we reconstruct $D_V/r_d$ by dividing with $r_{\mbox{\tiny d, fid}} = 147.49$ wherever applicable. We can rewrite \eqref{Dv} in terms of the reconstructed $D(z)$ and $H(z)$ from the CC data as,
\begin{equation}
	\left. D_V \right\vert_{\mbox{\tiny H}} = \left[\frac{c^2  D^2 (z) }{H_0^2} \frac{ c z}{H(z)}\right]^{\frac{1}{3}} = \frac{c}{H_0} \left[ \frac{D^2(z) z}{E(z)}\right]^{\frac{1}{3}}
\end{equation} Finally we minimize the $\chi^2$ function, \begin{equation} \label{chiBAO}
	\chi_{\mbox{\small BAO}}^2 = \sum \frac{\left( \left.\frac{D_V}{r_d}\right\vert_{\mbox{\tiny BAO}} - \left.\frac{D_V}{r_d}\right\vert_{\mbox{\tiny H}}\right)^2}{{\sigma_{\left. \frac{D_V}{r_d}\right\vert_{\mbox{\tiny BAO}}}^2} +{\sigma_{\left. \frac{D_V}{r_d}\right\vert_{\mbox{\tiny H}}}^2}}
\end{equation} and obtain the marginalized constraints on $r_d$ for a uniform prior assumption $r_d \in [130, 160]$. \\

Uncertainty in the parameters $M_B$, $\Omega_{m0}$ and $r_d$ are obtained by a Markov Chain Monte Carlo (MCMC) analysis. Here, we adopt a python implementation of the ensemble sampler for MCMC, the \texttt{emcee}\footnote{\url{https://github.com/dfm/emcee}}, introduced by \citet{emcee}. Finally, we plot the results using the \texttt{GetDist}\footnote{\url{https://github.com/cmbant/getdist}} module of python, developed by \citet{getdist}. Plots for the marginalized $M_B$, $\Omega_{m0}$ and $r_d$ constraints are shown in the Figures \ref{MB_plot}, \ref{Om_plot} and \ref{rd_plot}. The best fitting values are given in Table \ref{MOr_res}. \\

Another method for estimating the value of $r_d$ is from equation \eqref{rdrag_approx}. Using the marginalized constraints of $\Omega_{m0}$, and the reconstructed value of $H_0$ we evaluate $r_d$ along with it's 1$\sigma$ uncertainty by propagation of error. Moreover, we assumed the value $\Omega_{b0}h_0^2 = 0.022383$ from Planck 2018 \cite{plc}. Result for the calculated values of $r_d$ using the aforementioned procedure is shown in Table \ref{rdrag_tab}.\\

Thereafter, we plot the reconstructed dimensionless functions $D_L(z)$ [where $D_L = \frac{H_0 ~d_L}{c}$ is the normalized luminosity distance] and $\frac{D_v}{r_d}(z)$ in the redshift range $0 < z < 2$ as that of reconstructed $H(z)$, using Gaussian Process (GP) for different choices of the covariance function in Fig \ref{Dl_plot} and \ref{Dv_plot} respectively. The specific points (in the $H$, $D_L$ and $\frac{D_V}{r_d}$ plots) with error bars represent the observational data used in reconstruction. The hyperparameters ($\sigma_f, l$) are optimized by maximizing log marginal likelihood while performing the GP.\\

Finally, we perform the non-parametric reconstruction of $\eta$ directly using the reconstructed $H(z)$,  $d_L(z)$ and $D_V(z)$ following the relation,
\begin{equation}
	\eta(z) = \frac{d_L ~\sqrt{c ~z}}{D_V^{\frac{3}{2}}~ H^{\frac{1}{2}} ~(1+z)}.
\end{equation} Plots for the reconstructed $\eta(z)$ are shown in Fig \ref{eta_np_plot} and \ref{eta_np_plot2} considering different choices of the covariance function. In case of Fig \ref{eta_np_plot}, we make use the marginalized constraints on $r_d$ obtained via equation \eqref{chiBAO}. Fig \ref{eta_np_plot2} is plot considering the approximated value of $r_d$ derived from equation \eqref{rdrag_approx}. It appears so in Fig \ref{eta_np_plot2}, that the uncertainty does not increases significantly at higher redshift, as compared to Fig \ref{eta_np_plot}. However, comparing the Y-axes range of both the plots, it can be clearly seen that Fig \ref{eta_np_plot} is better constrained than that of \ref{eta_np_plot2}. This could be the effect of 1$\sigma$ uncertainty in $r_d$ which quite large in case of Fig \ref{eta_np_plot2} as compared to Fig \ref{eta_np_plot}, tabulated in Tables \ref{rdrag_tab} and \ref{MOr_res} respectively. The black solid line represents the best fit values of $\eta$. The shaded regions correspond to the 68 percent ($1\sigma$), 95 percent ($2\sigma$) and 99.7 percent ($3\sigma$) confidence levels (CL). Plots reveal that the reconstructed values of $\eta(z)$ in the low redshift range $0<z<1$ remains very close to unity. At higher redshift $z>1.5$, in case of the Mat\'{e}rn 5/2 and 7/2 covariance functions, CDDR is always allowed in 1$\sigma$ for Fig \ref{eta_np_plot2} and barely in case of Fig \ref{eta_np_plot}. Again, considering the Mat\'{e}rn 9/2 and Squared Exponential covariance functions, CDDR is almost allowed within the $1\sigma$ uncertainty in Fig \ref{eta_np_plot2} with increasing redshift $z>1.5$. Nonetheless, CDDR is always allowed within a 2$\sigma$ uncertainty, which indicate a non-violation of CDDR in the late time universe for all cases studied.

\section{Discussion}
\label{conclusion}

In the present study, viability of the cosmic distance duality relation is investigated by a non-parametric model-independent method. The distance modulus measurements of type Ia supernovae from the latest Pantheon sample, the cosmic chronometer measurements of the Hubble parameter, and the recent measurements of volume-averaged Baryon Acoustic Oscillation data are utilized in the analysis. Firstly, we perform a reconstruction of the luminosity distance $d_L$ from the Pantheon SN-Ia compilation. A reconstruction of the Hubble parameter $H(z)$ from the CC data, followed by another reconstruction of $D_V$ from the BAO has been carried out next using Gaussian Process. The analysis has been performed in the same domain of redshift for four choices of the covariance function, namely the Squared exponential, Mat\'{e}rn 9/2, Mat\'{e}rn 7/2 and Mat\'{e}rn 5/2 covariance. The angular distance $d_A$ are obtained via combining the reconstructed $H$ and $D_V$ from the CC and BAO data respectively.\\

Finally, the reconstructed functions viz. luminosity distance $d_L(z)$, Hubble parameter $H(z)$ and the volume-averaged distance $D_V$ have been combined to get the cosmic distance-duality ratio $\eta(z)$. The reconstructed $\eta$ curve is obtained directly from the data without assuming any prior functional form as ansatz. A common feature is that the best-fit curve for $\eta$, shown in Fig \ref{eta_np_plot} and \ref{eta_np_plot2}, is close to $\eta=1$ curve in the range $0<z<1$, and $\eta = 1$ is always included in 1$\sigma$ for the redshift range $0<z<1.5$. However with increasing $z$, i.e., $z>1.5$, in case of the Mat\'{e}rn 5/2 and 7/2 covariance functions, $\eta= 1$ is mostly included in 1$\sigma$ for Fig \ref{eta_np_plot2} and barely in 1$\sigma$ for Fig \ref{eta_np_plot}. In Fig \ref{eta_np_plot2}, $\eta=1$ value almost remains within the $1\sigma$ uncertainty for the Mat\'{e}rn 9/2 and Squared Exponential covariance functions. But, $\eta =1$ is always included in 2$\sigma$ for all choices of the covariance functions, in case of both Fig \ref{eta_np_plot} and \ref{eta_np_plot2}. It has already been mentioned that, this contrast in results arises due to the 1$\sigma$ uncertainty difference in $r_d$ values considered, while performing the reconstruction. As the BAO data points are mostly concentrated up to $z \simeq 1$, the reconstructed $\eta$ is well constrained up to $z\simeq 1$. The uncertainty in the $\eta(z)$ curves increase with increasing redshift. \\

It deserves mention that we obtained the marginalized constraints on $M_B$ by keeping the nuisance parameters $\alpha$, $\beta$,  colour, stretch and bias corrections fixed using the BBC framework, for the Pantheon compilation of SN-Ia. Fixing the value of $M_B$ from the global $\Lambda$CDM fits from \citet{pan1}, may result in inconsistencies as $M_B$ is degenerate with $H_0$. In case of the BAO $\frac{D_V}{r_d}$ data, the comoving sound horizon at photon drag epoch, $r_d$ is constrained considering a fiducial measure on $r_{\mbox{\tiny d, fid}}$ equal to $147.49$ wherever applicable. Again BAO measurement from WiggleZ in \citet{blake2012}, are cosmological parameter dependent, precisely $H_0$ and $\Omega_{m0}$. Therefore we have also derived the constraints on $\Omega_{m0}$ assuming a fiducial $\Lambda$CDM model. The uncertainties on these parameters are propagated properly for the evaluation of error in the distance measurements. As there may be correlations between these parameters $H_0$, $M_B$, $\Omega_{m0}$ and  $r_d$ keeping these values fixed may have a serious effect on the model-independent nature of reconstruction, if proper care is not taken.\\

The Gaussian process method has previously been used for a non-parametric reconstruction of $\eta$ in literature. \citet{nair3} utilized Union 2.1 SN-Ia data compilation, BAO data of SDSS, 6dF Galaxy Survey, WiggleZ and BOSS ($z=0.57$), and observational Hubble data compilation to reconstruct the luminosity distance, the angle-averaged distance  and the Hubble rate, using the GP regression technique in the redshift range $0.1 < z <0.73$. Our work is similar to the work by \citeauthor{nair3}, where the SN-Ia and BAO data were particularly used. But there are quite a few differences to list. We have used the latest updated Pantheon SN-Ia compilation which spans a redshift range $ z = 2.26$, instead of the Union 2.1 sample with available data points limited up to $z = 1.41$. Here, constraints have been obtained on a wider range of overlapping redshift $0 < z < 2$, and tighter constraints are obtained in the low redshift regime due to availability of more number of BAO data points. Another non-parametric reconstruction of the cosmic distance-duality relation by \citet{rana2017} using different dynamic and geometric properties of SGLs along with JLA SN-Ia observations, do not favour any deviation of CDDR and are in concordance with the standard value of unity within a 2$\sigma$ confidence region. In this case, the sole difference lies in terms of the data sets involved, and the redshift range considered for reconstruction. As for \citeauthor{rana2017} the reconstruction was mainly focused on the redshift range of $0<z<1$. \citet{zhou2019} reconstructed the distance-redshift relation from observations of the Dark Energy Survey SN-Ia with simulated fiducial $H(z)$ data, and obtained that, except for the very low redshift range $z<0.2$, there is no obvious deviation from the theoretical CDDR. The prime objective of work by \citeauthor{zhou2019} was to test the fidelity of Gaussian processes for cosmography where CDDR was reconstructed as a consistency check. The present work shows that the GP successfully reproduces the CDDR even at higher redshift.\\

The results obtained in the present analysis are totally in agreement with those from the existing literature, discussed above. We have extended the analysis to higher redshift. But due to the lack of observational data points at higher redshift, the uncertainty increases. We can conclude that all the recent studies of cosmic distance measurements support the theoretical CDDR at low redshift ($z<1.5$). Future higher redshift observations of BAO, SN-Ia and other observables would be able to provide tighter constraints on CDDR at higher redshift. Similar analysis with future observations would be useful to decide whether theoretically CDDR is equally valid at high redshift, or redshift dependent higher order correction terms are essentially required.

\section*{Acknowledgement}

The authors would like to thank Prof. Narayan Banerjee for useful discussions and suggestions. The authors would also like to acknowledge the anonymous referee for the important comments and constructive suggestions that led to a substantial improvement of the paper. AM acknowledges the financial support from the Science and Engineering Research Board (SERB), Department of Science and Technology, Government of India as a National Post-Doctoral Fellow (NPDF, File no. PDF/2018/001859).

%%%%%%%%%%%%%%%%%%%%%%%%%%%%%%%%%%%%%%%%%%%%%%%%%%
\section*{Data Availability}

Authors can confirm that all relevant source data are included in the article. The datasets generated during and/or analysed during the current study are available from the corresponding author on reasonable request.



\bibliographystyle{mnras}

\begin{thebibliography}{150}

\bibitem[\protect\citeauthoryear{Aghanim et al.}{2020}]{plc} Aghanim N. et al. [Planck Collaboration], 2020, \aap, 641, A6.

\bibitem[\protect\citeauthoryear{Alam et al.}{2017}]{alam2017} Alam S., Ata M., Bailey S., Beutler F. et al., 2017, \mnras, 470,
2617.

\bibitem[\protect\citeauthoryear{Alcaniz et al.}{2017}]{alcaniz2017} Alcaniz J. S. et al., 2017,  Fundam. Theor. Phys., 187, 11.

\bibitem[\protect\citeauthoryear{Anderson et al.}{2014}]{anderson2014} Anderson L. et al., 2014, \mnras, 441, 24.

\bibitem[\protect\citeauthoryear{Ata et al.}{2018}]{ata2018} Ata, M., Baumgarten, F., Bautista, J. et al. 2018, \mnras, 473, 4773.

\bibitem[\protect\citeauthoryear{Avgoustidis et al.}{2010}]{avgous2010} Avgoustidis A., Burrage C., Redondo J., Verde L., \& Jimenez R., 2010, \jcap, 10, 024.

\bibitem[\protect\citeauthoryear{Bassett \& Kunz}{2004a}]{bassett} Bassett B. A., Kunz M., 2004, \prd, 69, 101305.

\bibitem[\protect\citeauthoryear{Bassett \& Kunz}{2004b}]{bassett2} Bassett B. A., Kunz M., 2004, \apj, 607, 661.

\bibitem[\protect\citeauthoryear{Bautista et al.}{2017}]{bautista2017} Bautista J. E. et al., 2017, \aap, 603, A12.

\bibitem[\protect\citeauthoryear{Bautista et al.}{2018}]{bautista2018} Bautista, J. E., Vargas-Maga\~{n}a, M., Dawson, K. S. et al. 2018, \apj, 863, 110.

\bibitem[\protect\citeauthoryear{Betoule et al.}{2014}]{betoule2014} Betoule M. et al. [SDSS Collaboration], 2014, \aap, 568, A22.

\bibitem[\protect\citeauthoryear{Beutler et al.}{2011}]{beutler2011} Beutler, F., Blake, C., Colless, M. et al. 2011, \mnras, 416, 3017.

\bibitem[\protect\citeauthoryear{Blake et al.}{2012}]{blake2012} Blake C., Brough S., Colless M. et al., 2012, \mnras, 425, 405.

\bibitem[\protect\citeauthoryear{Blomqvist et al.}{2019}]{blomqvist} Blomqvist M. et al., 2019, \aap, 629, A86.

\bibitem[\protect\citeauthoryear{Bonamente et al.}{2006}]{bonamente2006} Bonamente M., Joy M. K., LaRoque S. J., Carlstrom J. E., Reese E. D., Dawson K. S., 2006, \apj, 647, 25.

\bibitem[\protect\citeauthoryear{Cardone et al.}{2012}]{cardone2012} Cardone V. F., Spiro S., Hook I., Scaramella, R., 2016, \prd, 85, 123510.


\bibitem[\protect\citeauthoryear{Carvalho et al.}{2016}]{carvalho2016} Carvalho G. C. et al., 2016, \prd, 93, 023530.

\bibitem[\protect\citeauthoryear{Carvalho et al.}{2017}]{carvalho2017} Carvalho G. C. et al., 2017, preprint, arXiv:1709.00271.

\bibitem[\protect\citeauthoryear{Cao et al.}{2015}]{cao2015} Cao S., Biesiada M., Gavazzi R., Pi\'{o}rkowska A., Zhu Z.-H., 2015, \apj, 806, 185.

\bibitem[\protect\citeauthoryear{Chen, Zhou \& Fu}{2015}]{chen2015} Chen Z., Zhou B., Fu X., 2015, Int. J. Theo. Phys., 55, 1229.

\bibitem[\protect\citeauthoryear{Corasaniti}{2006}]{corasaniti} Corasaniti P. S., 2006, \mnras, 372, 191.

\bibitem[\protect\citeauthoryear{de Carvalho et al.}{2018}]{carvalho2018} de Carvalho E. et al., 2018, \jcap, 04, 064.

\bibitem[\protect\citeauthoryear{de Carvalho et al.}{2020}]{carvalho2020} de Carvalho E. et al., 2020, \mnras, 492, 4469.

\bibitem[\protect\citeauthoryear{da Costa, Busti \& Holanda}{2015}]{costa2015} da Costa S. S., Busti V. C., Holanda R. F., 2015, \jcap, 10, 061.

\bibitem[\protect\citeauthoryear{de Sainte Agathe et al.}{2019}]{agathe} de Sainte Agathe V. et al., 2019, \aap, 629, A85.

\bibitem[\protect\citeauthoryear{Eisenstein \& Hu}{1998}]{rdrag_def} Eisenstein, D. J., Hu, W. 1998, \apj, 496, 605.

\bibitem[\protect\citeauthoryear{Ellis}{1971}]{ellis1971} Ellis G. F. R., 1971, General Relativity and Cosmology, Enrico Fermi Summer School Course XLVII, ed R. K. Sachs, New York Academic.

\bibitem[\protect\citeauthoryear{Ellis}{2007}]{ellis2007} Ellis G. F. R., 2007,  Gen. Rel. Grav., 39, 1047.

\bibitem[\protect\citeauthoryear{Ellis et al.}{2013}]{ellis2013} Ellis G. F. R., Poltis R., Uzan J. P., Weltman A., 2013, \prd, 87, 103530.

\bibitem[\protect\citeauthoryear{Etherington}{1933}]{etherin1993} Etherington I. M. H., 1933, The London, Edinburgh, and Dublin Philosophical Magazine and Journal of Science 15, 761.
	
\bibitem[\protect\citeauthoryear{Etherington}{2007}]{etherin2007} Etherington I. M. H., 2007, Gen. Rel. Grav., 39, 1055.

\bibitem[\protect\citeauthoryear{Filippis et al.}{2005}]{filipps2005} De Filippis E., Sereno M., Bautz M. W., Longo G., 2005, \apj, 625, 108.

%\bibitem[\protect\citeauthoryear{Font-Ribera et al.}{2014}]{font-ribera} Font-Ribera A. et al. [BOSS Collaboration], 2014, \jcap, 05, 027.

\bibitem[\protect\citeauthoryear{Foreman-Mackey et al.}{2013}]{emcee} Foreman-Mackey D., Hogg D. W., Lang D., Goodman J., 2013, Publ. Astron. Soc. Pac., 125, 306.

\bibitem[\protect\citeauthoryear{Fu, Zhou \& Chen}{2019}]{fu2019} Fu X., Zhou L., Chen J., 2019, \prd, 99, 083523.

\bibitem[\protect\citeauthoryear{Gil-Mar\'{i}n et al.}{2015}]{gilmarin2015} Gil-Mar\'{i}n, H., Percival, W. J., Cuesta, A. J. et al. 2016, \mnras, 460, 4210.

\bibitem[\protect\citeauthoryear{Gon\c{c}alves, Holanda \& Alcaniz}{2011}]{goncalves2011} Gon\c{c}alves R. S., Holanda R. F. L., Alcaniz J. S., 2011, \mnras, 420, L43.

\bibitem[\protect\citeauthoryear{Gon\c{c}alves et al.}{2015}]{gonclaves2014} Gon\c{c}alves R. S., Alcaniz J. S., Carvalho J. C., Holanda R. F. L., 2015, \prd, 91, 027302.

\bibitem[\protect\citeauthoryear{Gurvits}{1994}]{gurvits1994} Gurvits L. I., 1994, \apj, 425, 442.

\bibitem[\protect\citeauthoryear{Gurvits, Kellermann \& Frey}{1999}]{gurvits1999} Gurvits L. I., Kellermann K. I., Frey S., 1999, \aap, 342, 378.

\bibitem[\protect\citeauthoryear{Holanda, Lima \& Ribeiro}{2010}]{holanda2010} Holanda R. F. L., Lima J. A. S., Ribeiro M. B., 2010, \apj, 722, L233.

\bibitem[\protect\citeauthoryear{Holanda, Lima \& Ribeiro}{2011}]{holanda2011} Holanda R. F. L., Lima J. A. S., Ribeiro M. B., 2011, \aap, 528, L14.

\bibitem[\protect\citeauthoryear{Holanda, Gon\c{c}alves \& Alcaniz}{2012}]{holanda2012} Holanda R., Gon\c{c}alves R., Alcaniz J., 2012, \jcap, 12, 022.

\bibitem[\protect\citeauthoryear{Holanda \& Barros}{2016}]{holanda2016} Holanda  R., Barros K., 2016, \prd, 94, 023524.

\bibitem[\protect\citeauthoryear{Holanda, Busti \& Alcaniz}{2016}]{holanda2016b} Holanda R., Busti V., Alcaniz J., 2016, \jcap, 02, 054.

\bibitem[\protect\citeauthoryear{Holanda, Pereira \& da Costa}{2017}]{holanda2017} Holanda R., Pereira S., da Costa S. S., 2017, \prd, 95, 084006.

\bibitem[\protect\citeauthoryear{Holanda et al.}{2019}]{holanda2019} Holanda R., Cola\c{c}o L., Pereira S., Silva R., 2019, \jcap, 06, 008.

\bibitem[\protect\citeauthoryear{Hu \& Wang}{2018}]{hu2018} Hu J., Wang F. Y., 2018, \mnras, 477, 5064.

\bibitem[\protect\citeauthoryear{Jackson}{2004}]{jackson2004} Jackson J. C., 2004, \jcap, 11, 007.

\bibitem[\protect\citeauthoryear{Jhingan, Jain \& Nair}{2014}]{jhingan2014} Jhingan S., Jain D., Nair R., 2014, J. Phys. Conf. Ser. 484, 012035.

\bibitem[\protect\citeauthoryear{Kazin et al.}{2014}]{kazin2014} Kazin E. A. et al., 2014, \mnras, 441, 3524.

\bibitem[\protect\citeauthoryear{Kellermann}{1993}]{kellermann1993} Kellermann K. I., 1993, \nat, 361, 134.

\bibitem[\protect\citeauthoryear{Kessler \& Scolnic}{2017}]{beams} Kessler R., Scolnic D., 2017, \apj, 836, 56.

\bibitem[\protect\citeauthoryear{Lewis}{2019}]{getdist} Lewis A., 2019, preprint, arXiv:1910.13970 [astro-ph.IM].

\bibitem[\protect\citeauthoryear{Li \& Lin}{2018}]{li2018} Li X., Lin H.-N., 2018, \mnras, 474, 313.

\bibitem[\protect\citeauthoryear{Li, Wu \& Yu}{2011}]{li2011} Li Z., Wu  P., Yu H., 2011, \apj, 729, L14.

\bibitem[\protect\citeauthoryear{Liang et al.}{2013}]{liang2013} Liang N., Li Z., Wu P., Cao S., Liao K., Zhu Z.-H., 2013, \mnras, 436, 1017.

\bibitem[\protect\citeauthoryear{Liao et al.}{2016}]{liao2016} Liao K., Li Z., Cao S., Biesiada M., Zheng X., Zhu Z.-H., 2016, \apj, 822, 74.

\bibitem[\protect\citeauthoryear{Lima, Cunha \& Zanchin}{2011}]{lima2011} Lima J. A. S., Cunha J. V., Zanchin V. T., 2011, \apj, 742, L26.

\bibitem[\protect\citeauthoryear{Lin, Li \& Li}{2018}]{lin2018} Lin H.-N., Li M.-H., Li X., 2018, \mnras, 480, 3117.

\bibitem[\protect\citeauthoryear{Ma, Corasaniti \& Bassett}{2016}]{ma2016} Ma C., Corasaniti P. S., Bassett B. A., 2016, \mnras, 463, 1651.

\bibitem[\protect\citeauthoryear{Ma \& Corasaniti}{2018}]{ma2018} Ma C., Corasaniti P. S., 2018, \apj, 861, 124.

\bibitem[\protect\citeauthoryear{Max-Moerbeck et al.}{2018}]{max2014} Max-Moerbeck W., Richards J. L., Hovatta T., et al., 2014, \mnras, 445, 437.

\bibitem[\protect\citeauthoryear{Meng et al.}{2012}]{meng2012} Meng X.-L., Zhang T.-J., Zhan H., Wang X., 2012, \apj, 745, 98.

\bibitem[\protect\citeauthoryear{Moresco et al.}{2012}]{cc3} Moresco M., Cimatti A., Jimenez R., Pozzetti L., 2012, \jcap, 08, 006.

\bibitem[\protect\citeauthoryear{Moresco}{2015}]{cc4} Moresco M., 2015, \mnras, 450, L16.

\bibitem[\protect\citeauthoryear{Moresco et al.}{2016}]{cc5} Moresco M., Pozzetti L., Cimatti  A. et al., 2016, \jcap, 05, 014.

\bibitem[\protect\citeauthoryear{Mukherjee \& Banerjee}{2021}]{purba_jerk} Mukherjee P., Banerjee N., 2021, Euro. Phys. J. C, 81, 36.

\bibitem[\protect\citeauthoryear{Nair, Jhingan \& Jain}{2011}]{nair} Nair R., Jhingan S., Jain D., 2011, \jcap, 05, 023.

\bibitem[\protect\citeauthoryear{Nair, Jhingan \& Jain}{2012}]{nair2} Nair R., Jhingan S., Jain D., 2012, \jcap, 12, 028.

\bibitem[\protect\citeauthoryear{Nair, Jhingan \& Jain}{2015}]{nair3} Nair R., Jhingan S., Jain D., 2015, \prb, 745, 64.

\bibitem[\protect\citeauthoryear{Perlmutter et al.}{1999}]{perl1999} Perlmutter S. et al., 1998, \apj, 517, 565.

\bibitem[\protect\citeauthoryear{Rana et al.}{2017}]{rana2017} Rana A., Jain D., Mahajan S., Mukherjee A., Holanda R., 2017, \jcap, 07, 010.

\bibitem[\protect\citeauthoryear{Rana et al.}{2016}]{rana2016} Rana A., Jain D., Mahajan S., Mukherjee A., 2016, \jcap, 07, 026.

\bibitem[\protect\citeauthoryear{Ratsimbazafy et al.}{2017}]{cc6} Ratsimbazafy A. L., Loubser S. I., Crawford  S. M. et al., 2017, \mnras, 467, 3239.

\bibitem[\protect\citeauthoryear{Riess et al.}{1998}]{riess1998} Riess A. G. et al., 1998, \aj, 116, 1009.

\bibitem[\protect\citeauthoryear{Ross et al.}{2015}]{ross2015} Ross, A. J., Samushia, L., Howlett, C. et al. 2015, \mnras, 449, 835.

\bibitem[\protect\citeauthoryear{Ruan, Melia \& Zhang}{2018}]{ruan2018} Ruan C.-Z., Melia F., Zhang T.-J., 2018, \apj, 866, 31.

\bibitem[\protect\citeauthoryear{Samushia et al.}{2014}]{samushia2014} Samushia L., Reid B. A., White M. et al., 2014, \mnras, 439, 3504.

\bibitem[\protect\citeauthoryear{Scolnic et al.}{2018}]{pan1} Scolnic D. M. et al., 2018, \apj, 859, 101.

\bibitem[\protect\citeauthoryear{Seikel, Clarkson \& Smith}{2012}]{gapp} Seikel M., Clarkson C., Smith M., 2012, \jcap, 06, 036.

\bibitem[\protect\citeauthoryear{Simon et al.}{2005}]{cc0} Simon J., Verde L., Jimenez R., 2005, \prd 71, 123001.

\bibitem[\protect\citeauthoryear{Stern et al.}{2010}]{cc2} Stern D., Jimenez R., Verde L., Kamionkowski M., Stanford S. A., 2010, \jcap, 02, 008.

\bibitem[\protect\citeauthoryear{Suzuki et al.}{2012}]{union2.1} Suzuki N., Rubin D., Lidman C. et al., 2012, \apj, 746, 85.

\bibitem[\protect\citeauthoryear{Uzan, Aghanim \& Mellier}{2004}]{uzan2004} Uzan J.-P., Aghanim N., Mellier Y., 2004, \prd, 70, 083533.

\bibitem[\protect\citeauthoryear{Xu et al.}{2013}]{xu2013} Xu X., Cuesta A. J., Padmanabhan N., Eisenstein D. J., McBride C. K., 2013, \mnras, 431, 2834.

\bibitem[\protect\citeauthoryear{Xu \& Huang}{2020}]{xu2020} Xu B., Huang Q., 2020, Euro. Phys. J. Plus, 135, 447.

\bibitem[\protect\citeauthoryear{Zhang et al.}{2014}]{cc1} Zhang C., Zhang H., Yuan S., Zhang T.-J., Sun Y.-C., 2014, Res. Astron. Astrophys., 14, 1221.

\bibitem[\protect\citeauthoryear{Zheng et al.}{2019}]{zheng2019} Zheng J. et al., 2019, \mnras, 484, 442.

\bibitem[\protect\citeauthoryear{Zheng et al.}{2020}]{zheng2020} Zheng X., Liao K., Biesiada M., Cao S., Liu T.-H., Zhu Z.-H., 2020, \apj, 892, 103.

\bibitem[\protect\citeauthoryear{Zhou \& Li}{2019}]{zhou2019} Zhou H., Li Z. X., 2019, Chinese Phys. C, 43, 035103.



\end{thebibliography}

%%%%%%%%%%%%%%%%% APPENDICES %%%%%%%%%%%%%%%%%%%%%

%\appendix

% Don't change these lines
%\bsp	% typesetting comment
\label{lastpage}
\end{document}